# Structure-transport correlation reveals anisotropic charge transport in coupled PbS nanocrystal superlattices

Andre Maier[1,4], Dmitry Lapkin[2], Nastasia Mukharamova[2], Philipp Frech[1], Dameli Assalauova[2], Alexandr Ignatenko[2], Ruslan Khubbutdinov[2,5], Sergey Lazarev[2,6], Michael Sprung[2], Florian Laible[3,4], Ronny Löffler[4], Nicolas Previdi[1], Thomas Günkel[3], Monika Fleischer[3,4], Frank Schreiber[3,4], Ivan A. Vartanyants[2,5]∗, Marcus Scheele[1,4]∗

1. Institut für Physikalische und Theoretische Chemie, Universität Tübingen, Auf der Morgenstelle 18, D-72076 Tübingen, Germany
2. Deutsches Elektronen-Synchrotron DESY, Notkestraße 85, D-22607 Hamburg, Germany
3. Institut für Angewandte Physik, Universität Tübingen, Auf der Morgenstelle 10, D-72076 Tübingen, Germany
4. Center for Light-Matter Interaction, Sensors & Analytics LISA[+], Universität Tübingen, Auf der Morgenstelle 15, D-72076 Tübingen, Germany
5. National Research Nuclear University MEPhI (Moscow Engineering Physics Institute), Kashirskoe shosse 31, 115409 Moscow, Russia
6. National Research Tomsk Polytechnic University (TPU), pr. Lenina 30, 634050 Tomsk, Russia
∗. To whom correspondence should be addressed:
ivan.vartaniants@desy.de, marcus.scheele@uni-tuebingen.de

**Abstract**

Semiconductive nanocrystals (NCs) can be self-assembled into ordered superlattices (SLs) to create artificial solids with emerging collective properties.[1–3] Computational studies have predicted that properties such as electronic coupling or charge transport are determined not only by the individual NCs but also by the degree of their organization and structure.[4–7] However, experimental proof for a correlation between structure and charge transport in NC SLs is still pending. Here, we perform X-ray nano-diffraction and apply Angular X-ray Cross-Correlation Analysis (AXCCA)[8–10] to characterize the structures of coupled PbS NC SLs, which are directly correlated with the electronic properties of the same SL microdomains. We find strong evidence for the effect of SL crystallinity on charge transport and reveal anisotropic charge transport in highly ordered monocrystalline hexagonal close-packed PbS NC SLs, caused by the dominant effect of shortest interparticle distance. This implies that transport anisotropy should be a general feature of weakly coupled NC SLs.

**Main**

Previous experimental research on NC SLs has either focused solely on the process of self-organization and structural order[8,11–15] or, in separate studies, on charge transport and electronic properties.[16–21] In order to reveal potential transport anisotropy, a correlated investigation of charge transport and structural order on the same NC SL is required. This allows addressing a variety of fundamental questions. Are the electronic properties of NC SLs influenced by the SL type and orientation? Do polycrystalline and monocrystalline SLs differ in conductivity? What is the degree of transport anisotropy in NC SLs? Here, we address these questions by a direct correlation of the structural and electronic properties of SLs composed of electronically coupled PbS NCs.

As a model system we use oleic acid (OA) capped lead sulfide (PbS) NCs with a diameter of $5.8 \pm 0.5$ nm (Supplementary **Figure S1**), which are self-assembled and functionalized with the organic $\pi$-system Cu4APc (Cu-4,4',4'',4'''-tetraaminophthalocyanine) at the liquid-air interface.[22] This results in long-range ordered and highly conductive SLs.[23,24] By means of soft-lithographic microcontact printing,[25] we transfer stripes of PbS NC-Cu4APc SLs with a width ($W$) of roughly 4 µm onto trenches of ~1 µm length ($L$) between two gold contacts on X-ray transparent Kapton and Si/SiO$_x$ substrates (Supplementary **Figure S2**). This defines individually addressable microchannels with $L \approx 1$ µm, $W \approx 4$ µm, and thickness $h$ (**Figure 1a–e**). Since this area is comparable to the typical grain size of PbS NC SLs,[10] these microchannels enable transport measurements in single-crystalline PbS SLs.

In **Figure 2**, we display the charge transport characteristics of the microchannels as well as its dependence on the thickness of the SL and the probed area. The conductivity $\sigma$ is calculated as $\sigma = (G \cdot L)/(W \cdot h)$ for all individual microchannels from two-point probe conductance ($G$) measurements (**Figure 1e**, **Figure 2a**). Within the approximately two hundred individual microchannels measured, we observe electric conductivities in a wide range of values ($10^{-6}$–$10^{-3}$ S/m) (**Figure 2b**). This distribution correlates with the thickness of the SL (**Figure 2c**), which also varies by two orders of magnitude over the large number of microchannels analysed here. The correlation is non-linear with a maximum in $\sigma$ for thicknesses from 70 to 200 nm.

Using Si/SiO$_X$ as substrate, we performed field-effect transistor measurements of the PbS NC-Cu4APc SLs, revealing p-type behaviour, which agrees with our previous study[23] (Supplementary **Figure S3**). The microchannels show hole-mobilities up to $\mu \sim 10^{-4}$ cm$^2$ V$^{-1}$ s$^{-1}$.

We tested the effect of domain boundaries within the SL on electric transport on the same substrates measuring the geometry-normalized conductance of PbS NC SLs over large active channel areas of ~$10^4$ µm$^2$ (Supplementary **Figure S4**). As shown in **Figure 2d**, electric transport in this case is approximately two orders of magnitude less efficient than within the microchannels of ~4 µm$^2$, indicating the advantageous effect of the near single-crystalline channels present in the latter case (see below).

Further investigations of structural properties of the same microchannels on Kapton substrates using X-ray nano-diffraction in correlation with conductivity measurements are the focus of this study (**Figure 1e-f**, Supplementary **Figure S5**).

We determined the structural details of all microchannels by X-ray nano-diffraction (Methods/Supplementary **Figures S6–S9**). Using a nano-focused X-ray beam, we collected diffraction patterns at different positions in each channel (**Figure 1f**). Two typical small- and wide-angle (SAXS and WAXS, respectively) diffraction patterns from representative microchannels, averaged over all positions within these channels, are shown in **Figures 3a,g(b,h)**. For some of the microchannels we observe several orders of Bragg peaks in SAXS attributed to monocrystalline SLs (**Figure 3a**), whereas the rest of the channels demonstrate continuous Debye-Scherrer rings with low intensity modulations corresponding to polycrystalline SLs (**Figure 3g**). From the angular-averaged profiles, shown in **Figures 3c,i**, we revealed two dominant SL structures: a monocrystalline, random hexagonal close-packed (rhcp) lattice mainly oriented along the [0001]$_{SL}$, and a polycrystalline, body-centered cubic (bcc) lattice primarily oriented along the [110]$_{SL}$ (for SEM micrographs see Supplementary **Figure S13**). From the peak positions in SAXS, we estimated the unit cell parameters ($a_{rchp}$ and $a_{bcc}$) for each channel and corresponding nearest-neighbor distances (NND), that are $d_{NN} = a_{rhcp}$ for rhcp and $d_{NN} = \sqrt{3}/2 \cdot a_{bcc}$ for bcc. The averaged NNDs for all rhcp and bcc channels are 7.8 ± 0.4 and 6.9 ± 0.2 nm, respectively.

In WAXS (**Figures 3b,h**), we observe parts of three Debye-Scherrer rings corresponding to {111}$_{AL}$, {200}$_{AL}$, {220}$_{AL}$ reflections of the PbS atomic lattice (AL). From the single WAXS pattern analysis we found different degrees of angular disorder of NCs: roughly 24° for rhcp and 16° for bcc channels (Supplementary **Figure S9**).

To study the relative orientation of the NCs inside the SL, we applied AXCCA,[9] which is based on the analysis of the cross-correlation functions (CCFs), to the measured scattering data (Methods/Supplementary **Figures S10–S12**). We evaluated the CCFs for the SL and AL peaks for both rhcp and bcc structures. We found that in the rhcp monocrystalline channels (**Figure 3d**) the $[111]_{AL}$ and $[110]_{AL}$ directions of the NCs are collinear to the $[0001]_{SL}$ and $[2\bar{1}\bar{1}0]_{SL}$ directions, respectively (**Figure 3f**). In bcc polycrystalline channels (**Figure 3j**), all corresponding SL and AL directions are aligned (e.g. $\langle 100 \rangle_{SL}$ and $\langle 100 \rangle_{AL}$), as shown in **Figure 3l**. The similarity between the experimental CCFs and simulated CCFs for these structures confirms the obtained angular orientation of the NCs in the SL (**Figure 3d**,**e** and **3j**,**k**, respectively).

Upon correlating the X-ray with the electric transport measurements, we found that microchannels containing the polycrystalline bcc SLs exhibit higher conductivity than monocrystalline rhcp SLs over the entire range of thicknesses (**Figure 4a**). This can in part be understood in terms of the shorter NND which exponentially increases the hopping probability (**Figure 4b**).[4,17]

The microchannels exhibit strong characteristic Raman signals for Cu4APc (750 cm$^{-1}$ and 1,050–1,650 cm$^{-1}$) which vanish for probing areas outside the microchannels, verifying the specific functionalization of the NCs with the organic $\pi$-system (Supplementary **Figure S14**). We used the intensity of the two characteristic Raman bands to compare the relative density of Cu4APc molecules within different SLs. We found that polycrystalline bcc SLs with the smaller NND exhibit generally stronger Raman signals from Cu4APc than monocrystalline rhcp SLs with larger NND (**Figure 4c**,**d**, Supplementary **Figure S14**). This means that in monocrystalline rhcp SLs fewer native OA molecules have been exchanged by Cu4APc, resulting in larger interparticle distances, which adversely affects conductivity.

From **Figure 4b** one can identify several cases of monocrystalline rhcp SLs having conductivities as high as those of polycrystalline bcc SLs ($\sigma \sim 10^{-4}$–$10^{-3}$ S/m), although the NND is much larger. We consider this as supporting evidence that the degree of SL crystallinity (poly vs. mono) has a significant effect on the conductivity, which, in the present example, compensates the effect of the much larger interparticle distance.

The SLs with smaller interparticle distance exhibit stronger Raman signals from Cu4APc compared to larger SLs (Supplementary **Figure S14**), corroborating a correlation between interparticle distance and ligand exchange. In fact, the smallest lattice parameter of ~6.8 nm in **Figure 4b** corresponds to an interparticle distance of ~1 nm, which is approximately the length of one Cu4APc molecule or the minimal width of a fully exchanged ligand sphere. In contrast, residual OA leads to greater interparticle distances due to steric interactions of adjacent OA shells,[26] explaining the spread of the NNDs (**Figure 4b**, Supplementary **Figure S8**).

The occurrence of the two SL types (rhcp and bcc) found here may be related to the previously observed hcp-bcc transition for OA-capped PbS NC SLs upon tailored solvent evaporation.[27] Similarly, our polycrystalline bcc SLs are assembled from PbS NCs dispersed in hexane, whereas hexane-octane mixtures resulted in monocrystalline rhcp SLs. This invokes different solvent evaporation rates, which may lead to distinct SL unit cells.[15,26]

In view of the non-monotonic correlation between conductivity and SL thickness, we note that very thin NC films exhibit holes/microcracks, which are reduced with increasing thickness.[28] In contrast, the conductivity in thick films may be affected by a fringing electric field. The electric field is not homogeneous along the sample normal, and current flows mainly in the bottom layers close to the contacts. However, the conductivity is calculated over the entire channel where the full height is used.

We now turn to the key novelty of this work, the transport anisotropy, that is, the influence of the SL orientation with respect to the electric field on the electric conductivity.

For this, it is mandatory to account for the effect of SL thickness, incomplete ligand exchange and crystallinity and only compare SLs which are very similar in this regard. In doing so, we found strong evidence for a favoured angular direction of charge carrier hopping, indicating anisotropic charge transport within the SL. **Figures 5a**,**d** display exemplary SAXS patterns averaged over each microchannel of two monocrystalline rhcp SLs with identical structure, *i.e.* lattice parameter and thickness (Supplementary **Figure S15a**). They differ only in terms of the azimuthal orientation with respect to the applied electric field. We define the azimuthal angle $\alpha$ between the electric field vector $\boldsymbol{E}$ (which is oriented vertically due to horizontal electrode edges) and the nearest-neighbor direction $\boldsymbol{d}_{NN}$ (one of the $\langle 2\bar{1}\bar{1}0 \rangle_{SL}$ directions pointing to the nearest-neighbors). The angle $\alpha$ can vary from 0° to 30° for the sixfold in-plane symmetry. For $\alpha = 0°$, the $\boldsymbol{d}_{NN}$ direction is oriented parallel to the vector of electric field $\boldsymbol{E}$, whereas for $\alpha = 30°$, the angular (in-plane) offset between the vectors $\boldsymbol{E}$ and $\boldsymbol{d}_{NN}$ is maximized. Our key result is that for any two otherwise comparable channels, we observe higher conductivity for the respective channels with lower angle $\alpha$. The two extremes ($\alpha = 0°$ and $\alpha = 30°$) are shown in the corresponding real space SEM micrographs of the $(0001)_{SL}$ plane of two rhcp SLs in **Figure 5b**,**e**. The difference in conductivity between two otherwise identical SLs is 40–50%. A statistical investigation of other microchannels with monocrystalline rhcp SLs reveals similar $\alpha$-dependent conductivity differences (Supplementary **Figure S15**). This correlation between $\sigma$ and $\alpha$ indicates anisotropic charge transport, for which the direction of nearest neighbors is assumed to be the most efficient for transport.

In contrast to atomic crystals with transport anisotropy, which exhibit strong electronic coupling and ballistic transport (e.g. black phosphorus), the NC SLs studied here are in the weak coupling regime. This implies temperature-activated hopping as the predominant charge transport mechanism and invokes a strong dependence on the hopping distance.[4,17] Our results suggest that charge transport is most efficient if the applied electric field is iso-oriented with the nearest-neighbor direction $d_{NN}$ in the SL plane, since this leads to the shortest hopping distance (**Figure 5c**). Any other orientation (**Figure 5f**) results either in a larger hopping distance (straight arrow) or a deviation from the direction of the electric field together with an increased number of required jumps for electrons to travel the same distance (zig-zag path), which is detrimental to charge transport.

This implies that transport anisotropy should be a general feature of weakly coupled, monocrystalline NC SLs, originating from the dominant effect of the shortest interparticle distance. Accordingly, one could predict the favoured direction of charge transport within different SL types, such as simple cubic, face-centered cubic or body-centered cubic, being the ⟨100⟩, ⟨110⟩ or ⟨111⟩ SL directions, respectively. A similar charge transport anisotropy was computationally predicted for bcc and fcc SLs.[6]

Further, we note that the orientational order of the NCs observed here might be an additional source for anisotropic charge transport as different coupling strengths have been predicted along particular AL directions.[5,6] In the present case, the most efficient transport occurs if the $[110]_{AL}$ direction of all NCs is iso-oriented with the electric field.

A high degree of control provided over the SL type and orientation would enable the exploitation of such transport anisotropy also with more complex NC assemblies (e.g. binary NC SLs[29] or honeycomb structures[30]) for application in functional electronic devices with tailored transport anisotropy. Furthermore, these results constitute an important step towards the understanding of the intrinsic properties and fundamental limits of these fascinating new NC-based systems.

## Acknowledgements

This project has been funded by the DFG under Grant SCHE1905/3 and INST 37/829-1 FUGG, the latter for SEM measurements using a Hitachi SU 8030. Further funding was provided by the Helmholtz Associations Initiative, Networking Fund and the Russian Science Foundation grant HRSF-0002/18-41-06001 and the Tomsk Polytechnic University Competitiveness Enhancement Program grant (project number VIU-MNOL NK-214/2018). We acknowledge support and discussions with E. Weckert. T. Salditt's group is acknowledged for providing nano-focusing instrument (GINIX) support at the P10 beamline. The construction and operation of the GINIX setup is funded by the Georg-August Universität Göttingen and BMBF Verbundforschung "Struktur der Materie", projects 05KS7MGA, 05K10MGA, and 05K13MG4.

## Author contribution

A.M. performed the device fabrication, sample preparation, SEM, AFM, Raman and electrical measurements, analysed the results and wrote the manuscript. D.L. and N.M. performed the X-ray nano-diffraction experiment, analysed the results and wrote the manuscript. P.F. supported the sample preparation and electrical measurements. D.A., A.I., R.K., S.L., and M.Sp. performed and supported the X-ray nano-diffraction experiment. F.L., R.L., N.P., T.G. and M.F. supported the device fabrication. F.S., I.V. and M.Sch. conceived and supervised the project. All authors have given approval to the final version of the manuscript.

## Competing interests

The authors declare no competing interests.

## References

1. Murray, C. B., Kagan, C. R. & Bawendi, M. G. Self-organization of CdSe nanocrystallites into three-dimensional quantum dot superlattices. *Science* **270,** 1335–1338 (1995).

2. Alivisatos, A. P. Semiconductor clusters, nanocrystals, and quantum dots. *Science* **271,** 933–937 (1996).

3. Kagan, C. R. & Murray, C. B. Charge transport in strongly coupled quantum dot solids. *Nat. Nanotech.* **10,** 1013–1026 (2015).

4. Remacle, F. & Levine, R. D. Quantum dots as chemical building blocks. Elementary theoretical considerations. *ChemPhysChem* **2,** 20–36 (2001).

5. Yazdani, N. *et al.* Charge transport in semiconductors assembled from nanocrystals. Preprint at http://arxiv.org/pdf/1909.09739v1 (2019).

6. Kaushik, A. P., Lukose, B. & Clancy, P. The role of shape on electronic structure and charge transport in faceted PbSe nanocrystals. *ACS Nano* **8,** 2302–2317 (2014).

7. Liljeroth, P. *et al.* Variable orbital coupling in a two-dimensional quantum-dot solid probed on a local scale. *Phys. Rev. Lett.* **97,** 096803 (2006).

8. Zaluzhnyy, I. A. *et al.* Quantifying angular correlations between the atomic lattice and the superlattice of nanocrystals assembled with directional linking. *Nano Lett.* **17,** 3511–3517 (2017).

9. Zaluzhnyy, I. A. *et al.* Angular X-ray cross-correlation analysis (AXCCA): Basic concepts and recent applications to soft matter and nanomaterials. *Materials* **12,** 3464 (2019).

10. Mukharamova, N. *et al.* Revealing grain boundaries and defect formation in nanocrystal superlattices by nanodiffraction. *Small* **15,** 1904954 (2019).

11. Geuchies, J. J. *et al.* In situ study of the formation mechanism of two-dimensional superlattices from PbSe nanocrystals. *Nat. Mater.* **15,** 1248–1254 (2016).

12. Whitham, K., Smilgies, D.-M. & Hanrath, T. Entropic, enthalpic, and kinetic aspects of interfacial nanocrystal superlattice assembly and attachment. *Chem. Mater.* **30,** 54–63 (2018).

13. Weidman, M. C., Smilgies, D.-M. & Tisdale, W. A. Kinetics of the self-assembly of nanocrystal superlattices measured by real-time in situ X-ray scattering. *Nat. Mater.* **15,** 775–781 (2016).

14. Novák, J. *et al.* Site-specific ligand interactions favor the tetragonal distortion of PbS nanocrystal superlattices. *ACS Appl. Mater. Interfaces* **8,** 22526–22533 (2016).

15. Hanrath, T., Choi, J. J. & Smilgies, D.-M. Structure/processing relationships of highly ordered lead salt nanocrystal superlattices. *ACS Nano* **3,** 2975–2988 (2009).

16. Talapin, D. V. & Murray, C. B. PbSe nanocrystal solids for n- and p-channel thin film field-effect transistors. *Science* **310,** 86–89 (2005).

17. Liu, Y. *et al.* Dependence of carrier mobility on nanocrystal size and ligand length in PbSe nanocrystal solids. *Nano Lett.* **10,** 1960–1969 (2010).

18. Oh, S. J. *et al.* Stoichiometric control of lead chalcogenide nanocrystal solids to enhance their electronic and optoelectronic device performance. *ACS Nano* **7,** 2413–2421 (2013).

19. Evers, W. H. *et al.* High charge mobility in two-dimensional percolative networks of PbSe quantum dots connected by atomic bonds. *Nat. Commun.* **6,** 8195 (2015).

20. Whitham, K. *et al.* Charge transport and localization in atomically coherent quantum dot solids. *Nat. Mater.* **15,** 557–563 (2016).

21. Balazs, D. M. *et al.* Electron mobility of 24 $cm^2$ $V^{-1}$ $s^{-1}$ in PbSe colloidal-quantum-dot superlattices. *Adv. Mater.* **30,** 1802265 (2018).

22. Dong, A., Jiao, Y. & Milliron, D. J. Electronically coupled nanocrystal superlattice films by in situ ligand exchange at the liquid-air interface. *ACS Nano* **7,** 10978–10984 (2013).

23. André, A. *et al.* Structure, transport and photoconductance of PbS quantum dot monolayers functionalized with a copper phthalocyanine derivative. *Chem. Commun.* **53,** 1700–1703 (2017).

24. Maiti, S. *et al.* Understanding the formation of conductive mesocrystalline superlattices with cubic PbS nanocrystals at the liquid/air interface. *J. Phys. Chem. C* **123,** 1519–1526 (2019).

25. Qin, D., Xia, Y. & Whitesides, G. M. Soft lithography for micro- and nanoscale patterning. *Nat. Protoc.* **5,** 491–502 (2010).

26. Weidman, M. C., Nguyen, Q., Smilgies, D.-M. & Tisdale, W. A. Impact of size dispersity, ligand coverage, and ligand length on the structure of PbS nanocrystal superlattices. *Chem. Mater.* **30,** 807–816 (2018).

27. Lokteva, I., Koof, M., Walther, M., Grübel, G. & Lehmkühler, F. Monitoring nanocrystal self-assembly in real time using in situ small-angle X-ray scattering. *Small* **15,** 1900438 (2019).

28. Shaw, S. *et al.* Building materials from colloidal nanocrystal arrays. Preventing crack formation during ligand removal by controlling structure and solvation. *Adv. Mater.* **28,** 8892–8899 (2016).

29. Shevchenko, E. V., Talapin, D. V., Kotov, N. A., O'Brien, S. & Murray, C. B. Structural diversity in binary nanoparticle superlattices. *Nature* **439,** 55–59 (2006).

30. Boneschanscher, M. P. *et al.* Long-range orientation and atomic attachment of nanocrystals in 2D honeycomb superlattices. *Science* **344,** 1377–1380 (2014).

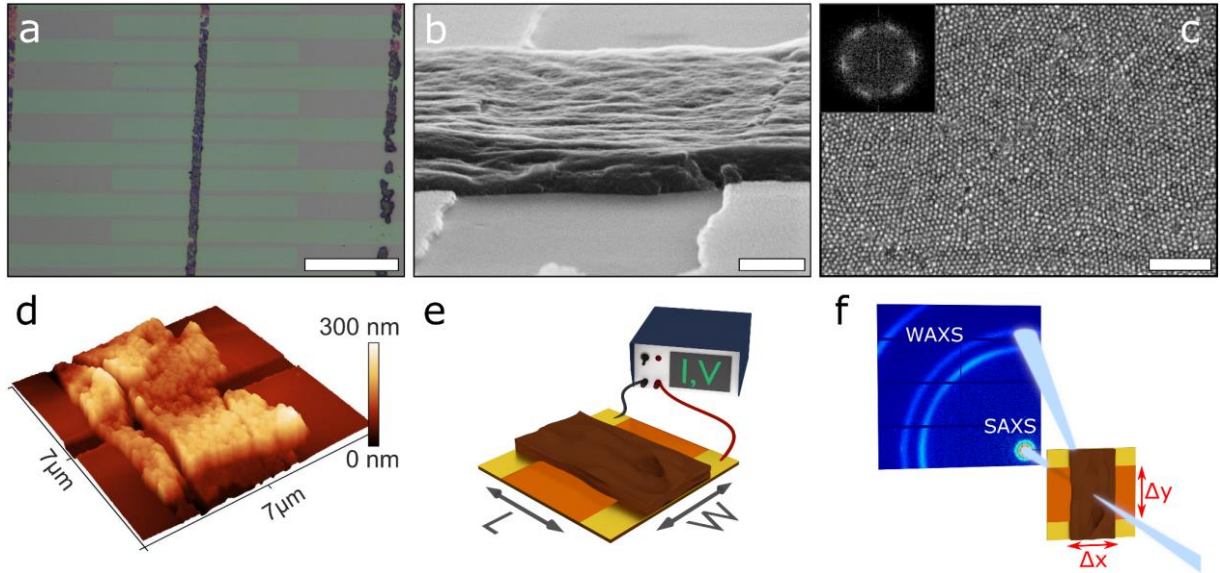

**Figure 1: Microchannels of PbS NC superlattices formed by microcontact printing to perform conductivity and X-ray nano-diffraction measurements.** (**a**) Optical micrograph of a set of electrode pairs (green). Orthogonal PbS NC stripes (brown), stamped via microcontact printing, connect adjacent electrodes to form microchannels, which can be individually addressed. Scale bar: 40 µm. (**b**) Scanning electron microscopy (SEM) micrograph in sideview (85° from normal) of a typical microchannel consisting of a ~200 nm thick PbS NC superlattice stripe across two Au electrodes. Scale bar: 300 nm. (**c**) High-resolution SEM micrograph showing self-assembled PbS NCs within a microchannel with near-range order, as indicated by the fast Fourier transform in the inset. Scale bar: 100 nm. (**d**) AFM micrograph of a typical microchannel on a Kapton substrate. (**e**) Scheme of a PbS NC domain (brown) bridging two electrodes (yellow) on a Kapton substrate (orange). A microchannel with length $L \approx 1\,\mu m$ and width $W \approx 4\,\mu m$ is formed to characterize the electronic properties of the PbS NC superlattice. (**f**) A nano-focused X-ray beam probes the structural order of the superlattice within the same microchannel by means of SAXS and WAXS. Spatial mapping is performed along Δx and Δy directions.

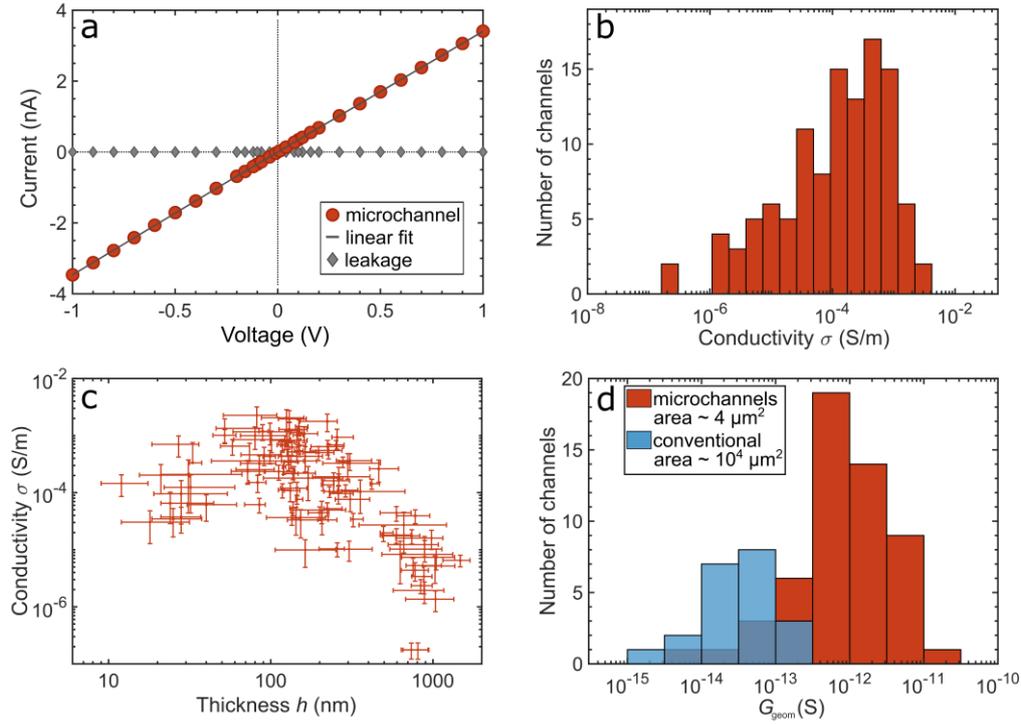

**Figure 2: Electrical transport measurements of PbS NC superlattice microchannels.** (**a**) Typical *I-V*-curve of PbS NC superlattice within a microchannel. Ohmic behaviour of the microchannel is observed (red) and fitting yields the conductance. The leak current through the dielectric substrate is negligible (grey). (**b**) Distribution of the electric conductivities of 200 individual microchannels. (**c**) Conductivity of the microchannels as a function of PbS NC superlattice thickness. Error bars represent the standard deviation of conductivity and the range of thickness determined by AFM, respectively. (**d**) Effect of channel area. Distribution of geometry-normalized conductance of conventional large area and microchannels, probing effective areas of ~$10^4$ µm$^2$ (blue) and ~4 µm$^2$ (red), respectively. Dark blue color corresponds to the overlap of the two distributions. All measured conductance values are normalized to the geometry of the channel (*L/W*).

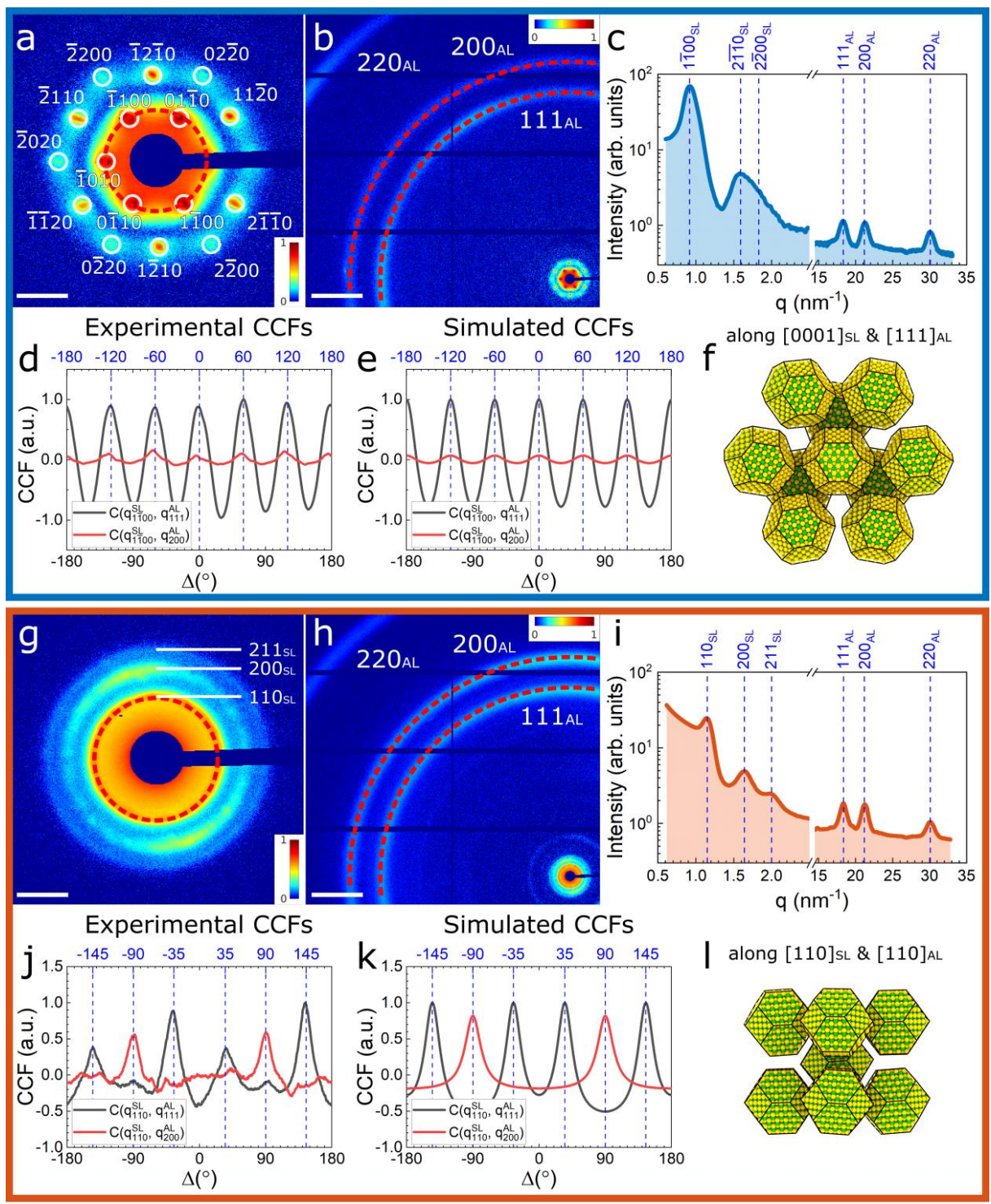

**Figure 3**: **Structural investigation of the SL and NCs AL.** Exemplary SAXS (**a,g**) and WAXS (**b,h**) patterns averaged over one microchannel for two typical cases: a monocrystalline rhcp superlattice oriented along the $[0001]_{SL}$ (**a,b**) and a polycrystalline bcc superlattice oriented along the $[110]_{SL}$ (**g,h**). The Bragg peaks are indexed accordingly. (**c,i**) Azimuthally averaged intensity profiles of SAXS (at $q < 2.5$ nm$^{-1}$) and WAXS (at $q > 15$ nm$^{-1}$) signals of the two superlattice types. (**d,j**) Averaged CCFs for the two superlattices, calculated for the first SAXS peaks ($\langle 1\bar{1}00\rangle_{SL}$ in the rhcp case (**d**) and $\langle 110\rangle_{SL}$

in the bcc case (**j**)) and the $\langle 111\rangle_{AL}$ or $\langle 200\rangle_{AL}$ WAXS peaks. (**e,k**) Simulated CCFs for the two models shown in (**f,l**). (**f,l**) Schematic drawing of the proposed superlattice structures: $[0001]_{SL}$-oriented rhcp superlattice of PbS NCs, where the NCs are aligned as shown in (**f**) and $[110]_{SL}$-oriented bcc superlattice of PbS NCs, where all the corresponding SL and AL directions are aligned (**l**). For clarity, ligand spheres are omitted. Scale bars in (**a,g**) and (**b,h**) correspond to 1 nm$^{-1}$ and 5 nm$^{-1}$, respectively.

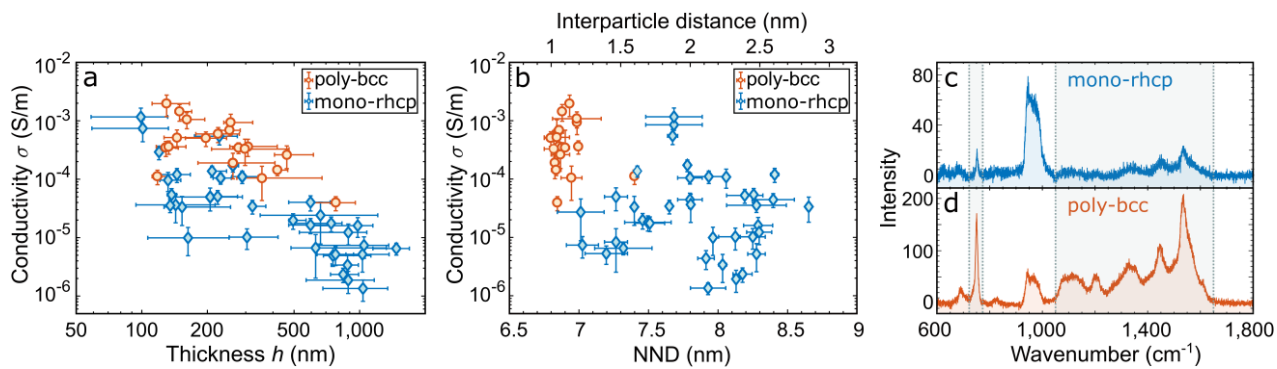

**Figure 4: Parameters for structure-transport correlations.** (**a**) Conductivity of individual microchannels as a function of superlattice thickness and superlattice type, indicated by the color code. (**b**) Conductivity of individual microchannels as a function of nearest-neighbour distance (NND). The superlattice type is indicated by the color code. (**c,d**) Typical Raman spectra of a monocrystalline rhcp (**c**) and a polycrystalline bcc (**d**) superlattice, featuring characteristic Cu4APc signals at 750 cm$^{-1}$ and 1,050–1,650 cm$^{-1}$ (highlighted regions). The polycrystalline bcc superlattices with smaller NND exhibits stronger Raman signal from Cu4APc, supporting the hypothesis of different degrees of ligand exchange. Additional information is given in Supplementary Figure S14. The signal at ~950 cm$^{-1}$ originates from the Si/SiO$_x$ of the substrate.

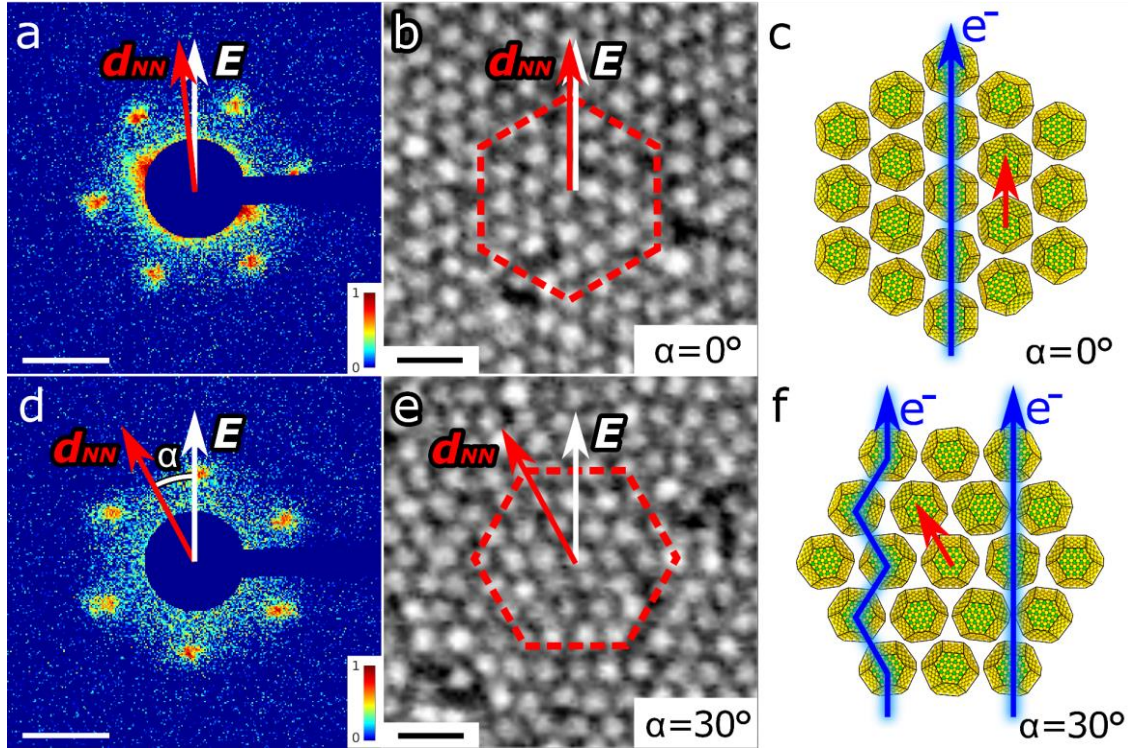

**Figure 5: Anisotropic charge transport in monocrystalline NC superlattices.** (**a,d**) Exemplary averaged SAXS diffraction patterns of comparable monocrystalline rhcp microchannels, oriented along [0001]$_{SL}$ crystallographic direction. The azimuthal orientation is defined by the relative angle $α$ between the vector of the electric field $E$ and the direction of NND $d_{NN}$. The superlattices with low values of $α$ feature 40–50% higher conductivity $σ$ than their counterparts with large $α$. The scale bar corresponds to 1 nm$^{-1}$. (**b,e**) Corresponding real space SEM micrographs of the NC superlattice oriented along the (0001)$_{SL}$ with $α = 0°$ and 30°. The hexagon indicates the orientation and hexagonal symmetry of the superlattice. $d_{NN}$ points along the alignment of the NCs (nearest neighbors). For $α = 0°$, the vector $d_{NN}$ is parallel to $E$, resulting in enhanced conductivity. The scale bar corresponds to 15 nm. (**c,f**) Schematic of the rhcp superlattice and the favoured hopping path for $α = 0°$ (blue arrow) along the $d_{NN}$ direction (red arrow) (**c**). For an in-plane offset ($α = 30°$), the larger hopping distance or the zig-zag path are detrimental to charge transport (**f**). Ligand spheres of NCs are omitted for clarity.

## Methods

**Device fabrication.** For the microchannel devices, we used photolithography techniques to pattern Au microelectrodes on either Si/SiO$_x$ wafer (200 nm SiO$_x$, n-doped Si, Siegert Wafer GmbH) or Kapton® polyimide membranes (DuPont™, 125 µm thickness). After exposure and development of the negative tone resist (ma-N 405, ma-D 331/S; purchased from Micro Resist Technology GmbH), 2.5 nm Ti as an adhesion layer and 8 nm Au were thermally evaporated under high vacuum conditions. Lift-off was performed in mr-Rem 660 (Micro Resist Technology GmbH), ultrasonic-assisted, to remove the resist and the metal layer on top, revealing electrodes with gaps of 0.7 to 1.7 µm (referred to as channel length $L$). Details on the device layout are given in the Supporting Information. Stamp masters based on silicon were fabricated by means of photolithography and pattern transfer via reactive ion etching. Defined trenches of 4 µm width and a periodicity of 80 µm were fabricated via anisotropic KOH etching of Si. Stamp masters were functionalized with F$_{13}$TCS (tridecafluoro-(1,1,2,2)-tetrahydrooctyl-trichlorosilane) as an anti-sticking layer. Degassed polydimethylsiloxane (PDMS, 10:1 ratio Sylgard 184 prepolymer and cross-linker) was poured onto the masters and cured at 150 °C overnight. Stamps were released from the master, cleaned by sonication in isopropanol and dried under pressurised nitrogen flow.

**Synthesis of PbS NCs.** OA-stabilised PbS NCs were synthesized according to the literature[31] and dispersed in hexane/octane (ratio from 1:0 to 0:1) at concentrations of about 5–10 µM. Applying sizing-curves to UV-Vis absorption spectra (in tetrachlorethylene) and SEM investigation yield a particle diameter of $5.8 \pm 0.5$ nm (size distribution of 8%), as indicated in Supplementary **Figure S1**.[32]

**Self-assembly and ligand exchange of PbS NC films.** PbS NCs were self-assembled via the liquid-air interface method, developed by Dong et al.[22,33] and modified as previously reported.[23,24,34,35] A certain volume (50–150 µl) of a PbS NC dispersion was injected on top of an acetonitrile subphase in a home-built Teflon chamber. The injection speed was controlled by a syringe pump. The evaporation rate of the dispersion solvent can be controlled by an adjustable lid/sealing. As the dispersion solvent evaporates, the NCs form a freely floating membrane. The Cu4APc ligand solution (Cu-4,4',4'',4'''-tetraaminophthalocyanine in dimethyl sulfoxide) was injected at the bottom of the liquid subphase. Cu4APc ligands diffuse through the liquid subphase to the NC membrane and replace the insulating native OA ligands over a duration of 4 h. The thickness of the floating film can be controlled to a certain degree (e.g. monolayer vs. thick film) by changing the NC dispersion volume, concentration and injection speed.

**Micro contact printing of PbS NC stripes.** A micropatterned PDMS stamp[25] was parallelly brought into contact with the free floating PbS NC membrane for 5 s. Excess liquid was removed from the stamp with a tissue. The coated stamp was slightly pressed onto the substrate with prepatterned Au-electrodes for 30 s. One half of the coated stamp was pressed onto a Kapton device, the other half onto a Si/SiO$_x$ device. Afterwards, the stamp was removed in a tilted manner. Vacuum-dried stamped substrates were placed on a spin coater and covered with acetone to remove unbound ligands. After 30 s, the solvent meniscus was removed by spinning at 20 rps for 30 s. This process was repeated twice. All preparation steps were performed in a nitrogen glovebox (level of O$_2$ < 0.5 ppm and H$_2$O = 0 ppm). Finally, the coated substrates were brought to ambient atmosphere for a defined time of 60 min and afterwards mounted into the probe station under nitrogen atmosphere. Individual channels consisting of an electrode pair and a connecting PbS-NC stripe were obtained. On a single device, up to 330 individual microchannels can be realised. A schematic drawing of the fabrication process is given in Supplementary **Figure S2**.

**Characterisation methods of PbS NC superlattice stripes.** Scanning electron microscopy (SEM) imaging of Si/SiO$_x$ devices was conducted with a HITACHI model SU8030 at 30 kV. Helium ion microscopy (HIM) imaging of Kapton devices was performed with a Zeiss ORION Nanofab at 30 kV. Using a flood gun, charge neutralisation on the sample can be achieved, to investigate insulating Kapton devices. For SEM and HIM sideview investigation of PbS-NC stripes, devices were analysed under a tilt angle of 85°. Atomic force microscopy (AFM) investigations were conducted with a Bruker MultiMode 8-HR. Raman spectroscopy measurements were performed with a confocal Raman spectrometer LabRAM HR800 (Horiba Jobin-Yvon) at a wavelength of 632.8 nm (He-Ne-laser) and a 100× objective.

**Electrical measurements.** All measurements of Si/SiO$_x$ and Kapton devices were performed at room temperature in a nitrogen flushed probe station (Lake Shore, CRX-6.5K). Individual Au-electrode pairs with a connected PbS-NC stripe were contacted with W-tips, connected to a source-meter-unit (Keithley, 2636B). A third electrode contacts the gate electrode (Si/SiO$_x$ device) or the rear of the dielectric (Kapton device). For two-point conductivity measurements of every microchannel, several voltage sweeps of ±1 V and ±200 mV were applied and the current detected (between two electrodes as well as leakage). For field-effect transistor (FET) measurements on Si/SiO$_x$, a source-drain voltage of $|V_{SD}| = 5$ V was applied and the current flow along the channel was modulated by applying a voltage sweep on the gate electrode (-40 V $\leq V_G \leq$ 40 V). Using the gradual channel approximation (Supporting Information, **Equation S1**), the field effect mobility $\mu$ of individual microchannels was calculated.

**X-ray nano-diffraction.** Nano-diffraction measurements were performed at Coherence beamline P10 of the PETRA III synchrotron source at DESY. An X-ray beam with wavelength $\lambda = 0.898$ nm ($E = 13.8$ keV) was focused down to a spot size of approximately $400 \times 400$ nm$^2$ (FWHM) with a focus depth of about 0.5 mm at the GINIX nano-diffraction endstation.[36] The two-dimensional detector EIGER X4M (Dectris) with $2070 \times 2167$ pixels and a pixel size of $75 \times 75$ μm$^2$ was positioned 370 mm downstream from the sample. The detector was aligned ~9 cm off-centre to allow simultaneous detection of small-angle (SAXS) and wide-angle (WAXS) X-ray scattering.

With an optical microscope, the most promising channels (based on electric transport measurements) were roughly localized. Precise localization of the individual channels was done using the WAXS scattering intensity of the Au {111} ($q_{111}^{Au} = 26.8$ nm$^{-1}$), as well as the PbS {111} and {200} reflections ($q_{111}^{PbS} = 18.3$ nm$^{-1}$, $q_{200}^{PbS} = 21.8$ nm$^{-1}$).

We then performed diffraction mapping of the entire coated area in each channel. Within this scanning region, diffraction patterns were collected on a raster grid with about 250 nm step size in both directions perpendicular to the incident beam (**Figure 1f**). The acquisition time was chosen to be 0.5 s in order to sustain a non-destructive regime of measurements. The chosen geometry allowed detecting the scattering signal from the NC superlattice as well as from the PbS AL simultaneously, but only a part of reciprocal space in WAXS was accessible. A sketch of the experimental scheme is shown in **Figure 1f**.

Using the nano-focused beam, it was possible to collect 100 to 200 diffraction patterns for each channel at different points within the channel. Integrating the WAXS intensity, we built diffraction maps of the microchannels (**Figure S6b**). A gap between two gold electrodes and the PbS NC superlattice across the microchannel are well observed. Noteworthily, the intensity modulation coincides with the AFM map of the same microchannel, shown in **Figure 1d**.

Averaging all individual diffraction patterns collected for a channel, we were able to study the average structure of the channel. From the azimuthally-averaged radial profiles we extracted the peak positions in SAXS and used them to calculate the superlattice unit cell parameter *a*. This analysis was performed for all measured channels.

**Angular X-ray Cross-Correlation Analysis.** To study relative orientation of the NCs inside the superlattice, we applied an Angular X-ray Cross-Correlation Analysis (AXCCA) approach[8–10] (see Supplementary for details). We calculated two-point cross-correlation functions (CCFs) for all channels, according to

$$C(q_{SL}, q_{AL}, \Delta) = \langle \tilde{I}(q_{SL}, \varphi) \tilde{I}(q_{AL}, \varphi + \Delta) \rangle_\varphi, \qquad (1)$$

where $\tilde{I}(q_{SL}, \varphi) = I(q_{SL}, \varphi) - \langle I(q_{SL}, \varphi) \rangle_\varphi$ and $I(q_{SL}, \varphi)$ is an intensity value taken at the point $(q_{SL}, \varphi)$ which are polar coordinates in the detector plane and $\langle \cdots \rangle_\varphi$ denotes averaging over all azimuthal $\varphi$ angles. In our analysis, momentum transfer values $q_{SL}$ correspond to SAXS peaks and $q_{AL}$ to WAXS peaks.

We calculated the CCFs for the first SAXS peaks ($\langle 1\bar{1}00 \rangle_{SL}$ in the rhcp case (**Figure 3d**) and $\langle 110 \rangle_{SL}$ in the bcc case (**Figure 3j**)) and the $\langle 111 \rangle_{AL}$ or $\langle 200 \rangle_{AL}$ WAXS peaks. From the peak positions at the CCFs we derived preferred angles between the corresponding SL and AL crystallographic directions. We proposed structural models satisfying the obtained angles. Based on the models, we simulated CCFs for each case (**Figure 3e,k**). Good agreement between the experimental and simulated CCFs verifies the proposed models.

**Methods References**

31. Weidman, M. C., Beck, M. E., Hoffman, R. S., Prins, F. & Tisdale, W. A. Monodisperse, air-stable PbS nanocrystals via precursor stoichiometry control. *ACS Nano* **8,** 6363–6371 (2014).

32. Moreels, I. *et al.* Size-dependent optical properties of colloidal PbS quantum dots. *ACS Nano* **3,** 3023–3030 (2009).

33. Dong, A. *et al.* Multiscale periodic assembly of striped nanocrystal superlattice films on a liquid surface. *Nano Lett.* **11,** 841–846 (2011).

34. André, A. *et al.* Toward conductive mesocrystalline assemblies: PbS nanocrystals cross-linked with tetrathiafulvalene dicarboxylate. *Chem. Mater.* **27,** 8105–8115 (2015).

35. Samadi Khoshkhoo, M., Maiti, S., Schreiber, F., Chassé, T. & Scheele, M. Surface functionalization with copper tetraaminophthalocyanine enables efficient charge transport in indium tin oxide nanocrystal thin films. *ACS Appl. Mater. Interfaces* **9,** 14197–14206 (2017).

36. Kalbfleisch, S. *et al.* The Göttingen holography endstation of beamline P10 at PETRA III/DESY. *AIP Conf. Proc.* **1365,** 96–99 (2011).

— Supplementary Information —

# Structure-transport correlation reveals anisotropic charge transport in coupled PbS nanocrystal superlattices

Andre Maier[1,4], Dmitry Lapkin[2], Nastasia Mukharamova[2], Philipp Frech[1], Dameli Assalauova[2], Alexandr Ignatenko[2], Ruslan Khubbutdinov[2,5], Sergey Lazarev[2,6], Michael Sprung[2], Florian Laible[3,4], Ronny Löffler[4], Nicolas Previdi[1], Thomas Günkel[3], Monika Fleischer[3,4], Frank Schreiber[3,4], Ivan A. Vartanyants[2,5]∗, Marcus Scheele[1,4]∗

1. Institut für Physikalische und Theoretische Chemie, Universität Tübingen, Auf der Morgenstelle 18, D-72076 Tübingen, Germany
2. Deutsches Elektronen-Synchrotron DESY, Notkestraße 85, D-22607 Hamburg, Germany
3. Institut für Angewandte Physik, Universität Tübingen, Auf der Morgenstelle 10, D-72076 Tübingen, Germany
4. Center for Light-Matter Interaction, Sensors & Analytics LISA+, Universität Tübingen, Auf der Morgenstelle 15, D-72076 Tübingen, Germany
5. National Research Nuclear University MEPhI (Moscow Engineering Physics Institute), Kashirskoe shosse 31, 115409 Moscow, Russia
6. National Research Tomsk Polytechnic University (TPU), pr. Lenina 30, 634050 Tomsk, Russia
∗. To whom correspondence should be addressed:

ivan.vartaniants@desy.de, marcus.scheele@uni-tuebingen.de

**Figure S1** displays the size investigation of oleic acid stabilized PbS nanocrystals (NC). From absorption spectra and SEM investigation, a diameter of 5.8 ± 0.5 nm is obtained.

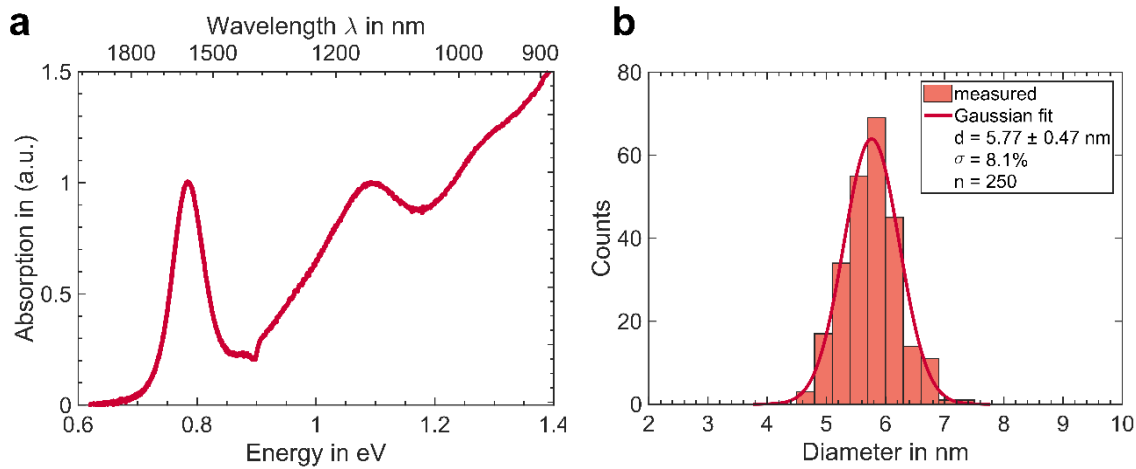

**Figure S1: PbS nanocrystal (NC) size analysis.** (**a**) Absorption spectrum of native oleic acid stabilized PbS NCs in tetrachlorethylene. The first excitonic transition is observed at 787 meV (1580 nm). (**b**) Distribution of the diameter of 250 PbS NCs, measured by SEM. Gaussian fit reveals a diameter of 5.8 ± 0.5 nm (size distribution of 8%).

**Figure S2** shows a schematic of the fabrication process of PbS NC superlattices. The NCs are self-assembled and functionalized with Cu4APc (Cu-4,4',4'',4'''-tetraaminophthalocyanine) at the liquid-air interface (**Figure S2a–c**) and stripes of the superlattices are transferred onto pre-patterned electrodes to form individually contactable microchannels by means of microcontact printing (**Figure S2d–g).**

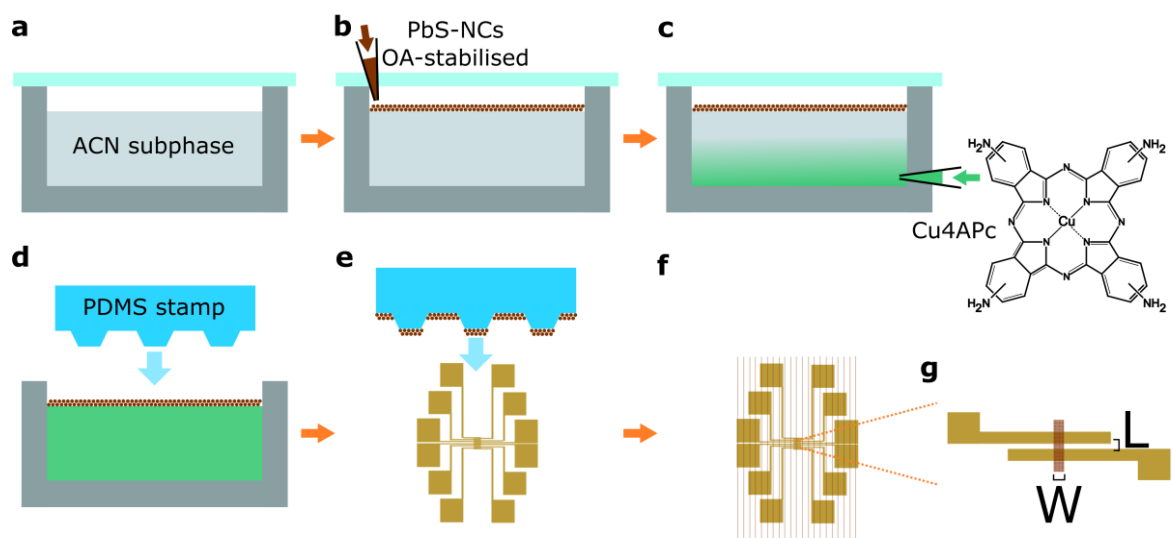

**Figure S2: Schematic drawing of the fabrication process of PbS NC superlattices and micro contact printing to form microchannels.** (**a**) A home-built Teflon chamber with an area of 1×1 cm$^2$ is filled with acetonitrile (ACN) as a subphase and covered with a glass slide. (**b**) A certain volume (50–150 µl) of a PbS NC dispersion (d = 5.8 nm, hexane/octane as solvent) is injected on top of the subphase. The injection speed is controlled by a syringe pump. The evaporation rate of the dispersion solvent can be controlled by an adjustable sealing. As the dispersion solvent evaporates, the NCs form a freely floating membrane. (**c**) The ligand solution (Cu4APc in dimethyl sulfoxid, DMSO) is injected into the bottom of the liquid subphase. Cu4APc ligands diffuse through the liquid subphase to the NC membrane and replace the insulating native oleic acid ligands over a duration of 4 h. (**d**) A micropatterned polydimethylsiloxane (PDMS) stamp is parallelly brought into contact with the free floating PbS NC membrane for 5 s. Excess liquid is removed from the stamp with a tissue and the stamp is dried for a few minutes. (**e**) The coated stamp is slightly pressed onto the substrate (Si/SiO$_x$ or Kapton devices) containing prepatterned Au-electrodes for 30 s (microcontact printing). This results in stripes of self-assembled PbS NC superlattices on electrode structures, as shown in (**f**). (**g**) This electrode-structure contains 11 individual channels with length *L* and width *W* of about 1 µm and 4 µm, respectively. Up to 30 electrode-structures are present on one device (330 individual microchannels).

**Layout of microchannel devices**

We developed microchannel devices, where individual electrode pairs with an overlap of 80 µm form channels, which can be addressed individually. Up to 330 channels per device could be realised. By means of microcontact printing,[1,2] we transfer periodic stripes of self-assembled superlattices with widths of 4 µm and a periodicity of 80 µm using prepatterned polydimethylsiloxane (PDMS) stamps. Thus, most of the substrate area remains uncoated. The microchannels are formed by pairs of overlapping electrodes, which are connected by one perpendicularly printed stripe of PbS-NC superlattice. The electrode thickness was chosen to be ~10 nm in order to avoid breaking of PbS-NC stripes at the edges. Due to the channel geometry, an entirely homogeneous electric field is established within the channel and the direction of the electric field vector is well-known.

**Field-effect transistor measurements on microchannels**

We conducted field-effect transistor (FET) measurements on microchannels. **Figure S3a** shows a typical transconductance curve, indicating p-type behavior. Using the gradual channel approximation (**Equation S1**), FET hole mobilities can be calculated, as indicated in **Figure S3b**. The p-type behavior and hole-mobilities are in line with previous studies.[3] The gradual channel approximation for FET characterisation is given in **Equation S1**.[4]

$$\mu = \frac{\partial I_{\mathrm{SD}}}{\partial V_{\mathrm{G}}} \frac{L}{W} \frac{t_{ox}}{\varepsilon_0 \varepsilon_{\mathrm{r}} V_{SD}} \qquad (S1)$$

Here, $\frac{\partial I_{\mathrm{SD}}}{\partial V_{\mathrm{G}}}$ is the derivation of $I_{\mathrm{SD}}$ in the transconductance curve, $V_{\mathrm{SD}}$ the applied source-drain voltage, $\varepsilon_0 \varepsilon_{\mathrm{r}}$ and $t_{\mathrm{ox}}$ the permittivity and the thickness of the dielectric $SiO_x$ layer, respectively. While the geometry of our microchannels is not ideal for typical FET measurements, this approach is sufficient for a qualitative comparison of different microchannels.

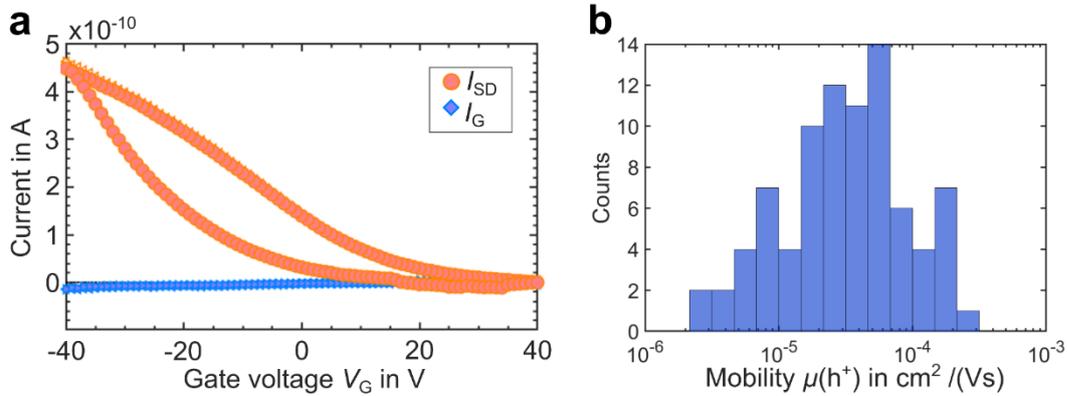

**Figure S3: FET measurements on microchannels.** (**a**) Transconductance curve of a PbS NC stripe on a Si/SiO$_x$ device. The source-drain current $I_{SD}$ can be modulated by the applied gate voltage $V_G$. The source drain voltage is set to $V_{SD}$ = 5 V. The leak current $I_G$ through the dielectric substrate is negligible. The channels show p-type behavior and the hole mobility can be calculated. (**b**) Distribution of field-effect hole mobilities $\mu(h^+)$ of individual microchannels. A log-normal distribution and a spread over 2 orders of magnitude is observed.

**Comparison of microchannels and state-of-the-art channels**

In conventional state-of-the-art electrode devices, interdigitated electrodes probe areas of approximately 1–20 × 10$^4$ µm$^2$ ($L$ ranging from 2.5 µm to 20 µm and $W \leq$ 1 cm). Typically, different domains are connected by the electrodes after coating (ranging from monolayer to several hundred nm), as exemplarily shown in **Figure S4a**. We normalized the conductance of 21 conventional channels and 54 microchannels (**Figure S4b**) of different thicknesses to the geometry ($G_{geom} = G \times [L/W]$). For huge conventional electrode devices, the conductivity cannot be calculated due to nonuniform thicknesses. The distributions (**Figure S4c**) can clearly be separated and the normalized conductance values of microchannels exceed those of conventional electrode devices.

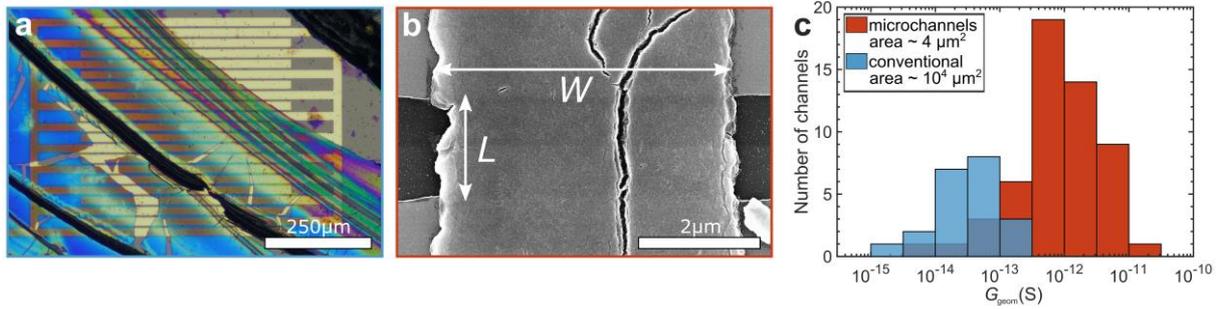

**Figure S4**: Comparison of conventional and microchannels. (**a**) Optical micrograph of a typical conventional state-of-the-art device of interdigitated electrodes with $L = 2.5–20$ µm and $W \leq 1$ cm. Active areas of $1–20 \times 10^4$ µm$^2$ are probed. (**b**) SEM micrograph of a typical microchannel with $L \sim 1–1.5$ µm and $W \sim 4$ µm. Active areas of only few µm$^2$ are probed. (**c**) Distribution of geometry-normalized conductance of conventional and microchannels (blue and red, respectively). Here, measured conductance values are normalized to the geometry of the channels ($L/W$).

### Microchannels on X-ray transparent Kapton devices

A typical Kapton device with 330 microchannels is shown in **Figure S5**. Kapton foil (polyimide, DuPont$^{TM}$) with a thickness of 125 µm was used as an X-ray transparent substrate. At this thickness, the Kapton foil is durable enough to warrant robust electric contacting and sufficiently X-ray transparent to enable scattering experiments. Further, it is robust enough for the photolithographic electrode fabrication process and allows fabrication of large-scaled devices ($15 \times 15$ mm$^2$). This allows performing X-ray diffraction on the entire substrate. The conductance of microchannels on Kapton devices can be determined and different channels exhibit different conductance with a large spread, as indicated in **Figure S5f**. We verified that the electronic transport measurements are not significantly influenced by the substrate (Si/SiO$_x$ or Kapton) itself.

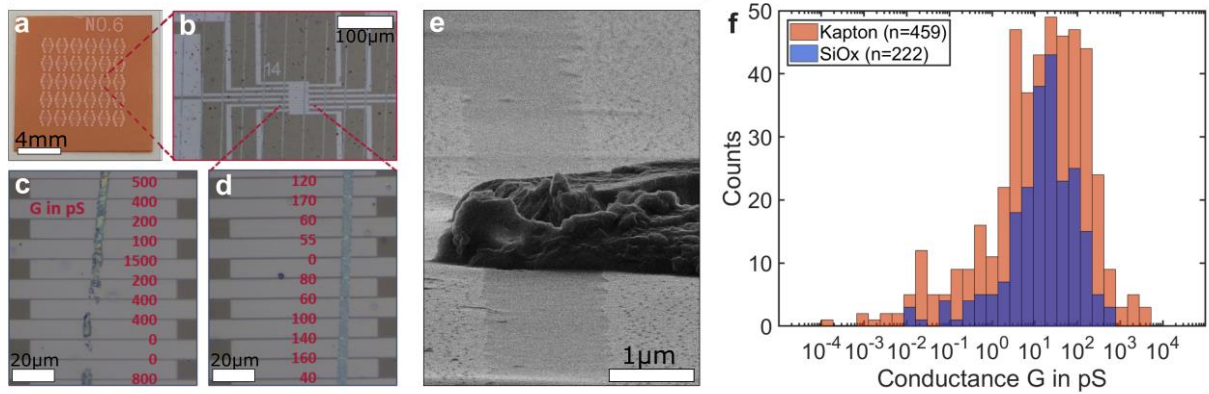

**Figure S5: Microchannels on X-ray transparent devices**. (**a**) Photograph of a Kapton device with 30 electrode structures. (**b**) Optical micrograph of typical electrode set with PbS-NC stripes across. (**c,d**) Optical micrographs of eleven individual microchannels each, which can be individually addressed and show different conductance values (indicated by values in red, *G* in pS). (**e**) Helium-ion microscopy micrograph showing side view of a typical microchannel (tilt angle of 85°). (**f**) Distribution of conductance values of *n* microchannels on Kapton and Si/SiO$_x$ devices.

### X-ray nano-diffraction signal detection and sample requirements

We note that acceptable signal-to-noise ratios during X-ray scattering were obtained only for thicker samples (> 100 nm). **Figure S6a** displays the superlattice thickness required to obtain the XRD-signals. Structural properties can only be investigated of superlattices with a minimum thickness of 100–200 nm. For microchannels with stripes of the required thickness, diffraction peaks in the SAXS region can clearly be identified. Hence, the PbS NCs within the stripes are highly ordered. Mapping areas of interest using the WAXS scattering intensity (the {111} Au ($q_{111}^{Au}$ = 26.8 nm$^{-1}$), {111} and {200} PbS reflections ($q_{111}^{PbS}$ = 18.3 nm$^{-1}$, $q_{200}^{PbS}$ = 21.8 nm$^{-1}$,)) allows to precisely localize the individual channels and the PbS NC superlattice within the latter. **Figure S6b** shows a typical diffraction map of a PbS NC superlattice in a microchannel. Every pixel corresponds to a single diffraction pattern. The horizontal electrode gap and the PbS NC superlattices across can clearly be identified. The intensity modulation coincides with the AFM map of the same microchannel, shown in **Figure 1d**.

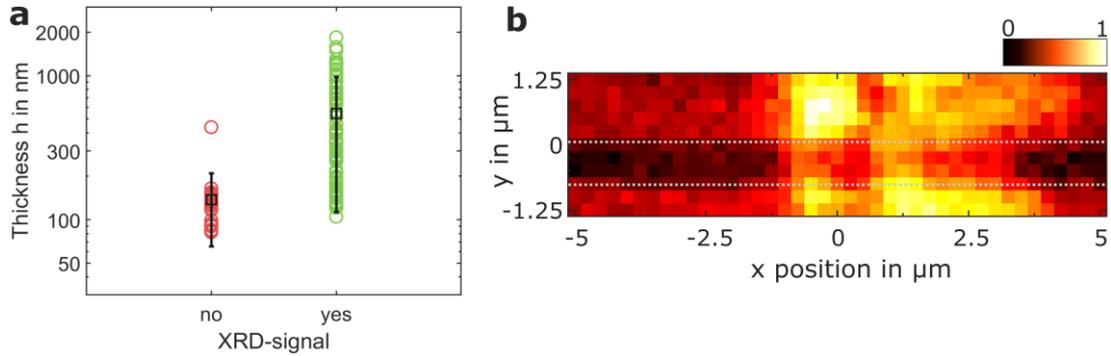

**Figure S6: X-ray nano-diffraction signal detection.** (**a**) Thickness of microchannel superlattice and corresponding XRD-signal. The mean thickness (± standard deviation) of superlattices with and without XRD-signal is 140 ± 82 nm and 545 ± 432 nm, respectively. (**b**) Typical intensity map of a PbS NC superlattice in a microchannel with pixel sizes of $250 \times 250$ nm$^2$. Every pixel corresponds to a single diffraction pattern. Averaging all diffraction patterns of a single microchannels allows to fully characterize the superlattice within. The color code indicates the XRD signal intensity. The diffraction map coincides with the AFM map of the same microchannel (**Figure 1d**).

**SAXS analysis for superlattice structure determination**

Analyzing averaged diffraction patterns for all measured channels, we found two groups among them. The first group of channels showed monocrystalline SAXS diffraction patterns (an example is shown in **Figure S7a**) and the second one showed Debye-Scherrer rings with relatively low angular intensity modulation in SAXS (**Figure S7d**). The monocrystalline patterns are of 6-fold symmetry and contain the Bragg peaks at $q_1$, $q_2 = \sqrt{3} \cdot q_1$ and $q_3 = 2 \cdot q_1$ (see **Figure S7c** for the radial profile), which can be attributed to a $[0001]_{SL}$-oriented random hexagonal close-packed structure (rhcp) superlattice (see details below). The presence of the Bragg peaks at $q_1$, $q_2 = \sqrt{2} \cdot q_1$ and $q_3 = \sqrt{3} \cdot q_1$ for the polycrystalline channels (see **Figure S7f** for the radial profile) is a clear evidence of a bcc superlattice structure. A single diffraction pattern for the polycrystalline channels contains the Bragg peaks from several grains with different orientations (see **Figure S7e**), thus the grain size is smaller than the beam size ($< 400 \times 400$ nm$^2$).

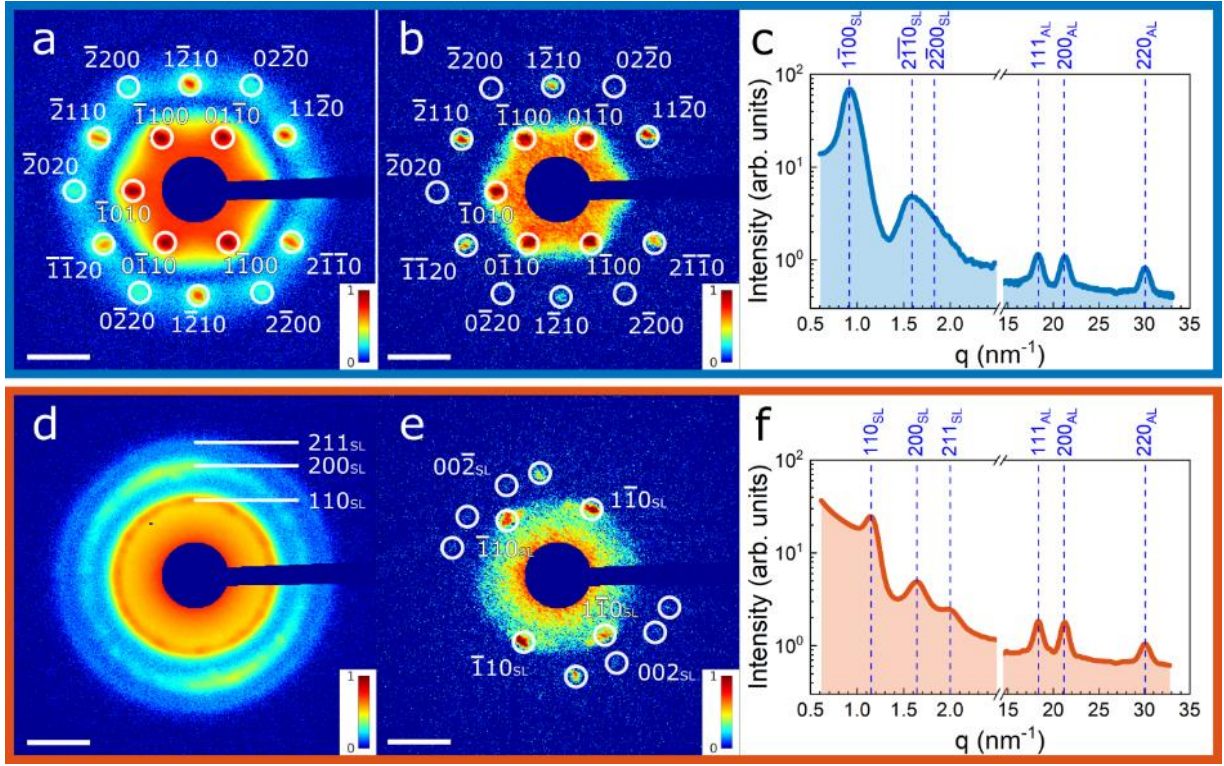

**Figure S7: Superlattice structure determination** for (**a–c**) mono- and (**d–f**) polycrystalline channels. (**a,d**) Averaged SAXS patterns; (**b,e**) examples of single SAXS patterns; (**c,f**) average radial profiles. Scale bars correspond to 1 nm$^{-1}$.

The obtained q-values were utilized to calculate the nearest-neighbor (center-to-center) distance between adjacent NCs in channels with both types of lattices. The distances are $6.9 \pm 0.2$ nm and $7.8 \pm 0.4$ nm for the poly- and monocrystalline channels, respectively (by fitting with normal distribution, see **Figure S8**). We use nearest-neighbor distances instead of unit-cell sizes in order to allow for a direct comparison between different superlattice types. The interparticle distances are calculated by subtracting the mean NC diameter from the nearest-neighbor distance (NND).

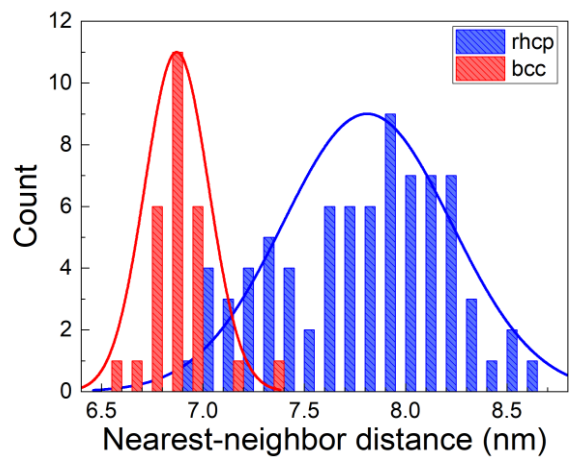

**Figure S8: Distribution of NND values for two types of channels.** Solid lines are fits by normal distribution.

**Superlattice structure of the monocrystalline channels**

The outline of our experiment (transmission geometry in one direction normal to the substrate) makes analysis of the monocrystalline samples quite complicated. The 6-fold patterns observed for the monocrystalline channels (**Figure S9a**) can be attributed to the hexagonal hcp lattice as well as to cubic bcc and fcc lattices. Indeed, the peak positions with respect to the first peak at $q_1$, $q_2 = \sqrt{3} \cdot q_1$ and $q_3 = 2 \cdot q_1$ correspond to all of them. NNDs are $d_{c-c} = 2\pi/q_1 \cdot (2/\sqrt{3})$, $2\pi/q_1 \cdot 2$ and $2\pi/q_1 \cdot (\sqrt{6}/2)$ for the hcp, fcc and bcc lattices, respectively. Using the mean $q_1$ value, one can obtain the NNDs of 7.8, 13.5 and 8.3 nm. In comparison to the size of the used NCs (~6.0 nm), the value obtained for the fcc seems to be unreasonable and we excluded this structure from our consideration.

The decision between hcp and bcc cannot be made based only on the diffraction patterns. First of all, we assumed the superlattice to have the same structure as the polycrystalline channels – a bcc superlattice with the NCs aligned with all superlattice directions. But in this case, the results of the cross-correlation analysis (see below, **Figure S11c**) are not consistent with the experimentally observed data. The NCs can be rotated inside the superlattice around the beam direction by 30º, but it would break the symmetry of the lattice. A possible arrangement of NCs in an hcp lattice is shown in **Figure S11d** and gives a cross-correlation function consistent with the experimental one (**Figure S11c**). This lattice has a higher symmetry than the proposed bcc lattice with rotated NCs. Also, the comparably big NND supposes a more sphere-like shape of the NCs covered with organic shell. A thicker shell makes interactions between adjacent NCs more isotropic. It is confirmed by the study of WAXS reflections from the NC ALs, that the angular disorder of NCs is higher for the monocrystalline channels than for the polycrystalline ones (see details below). On an average WAXS pattern of a monocrystalline channel typically only Debye-Scherrer rings are visible (**Figure S9a**). In this case, sphere-like particles with anisotropic interactions are likely to form close-packed structures like hcp. Thus, we assume the monocrystalline channels having an hcp structure. However, the close-packed structures are prone to form stacking faults leading to alternation of hcp and fcc structures. In our geometry (scattering along $[0001]_{SL}$ direction), we are not able to reveal fractions of both motifs. Thus, the correct description for the structure would be "random hcp (rhcp) lattice". We used this description throughout the manuscript, using the hcp-like indexing of the Bragg peaks (**Figure S7a,b**).

**WAXS analysis for NC alignment determination**

Analyzing the WAXS patterns from different types of channels, we noticed a drastic difference. The single patterns of the monocrystalline rhcp channels contain continuous Debye-Scherrer rings with low intensity modulations (**Figure S9a**), whereas for polycrystalline bcc channels, we found relatively sharp Bragg peaks of PbS atomic lattice (AL) reflections (**Figure S9d**). The difference in the intensities between the $111_{AL}$ and $200_{AL}$ reflections is caused by different out-of-plane NCs orientations with respect to the incident beam (the substrate).

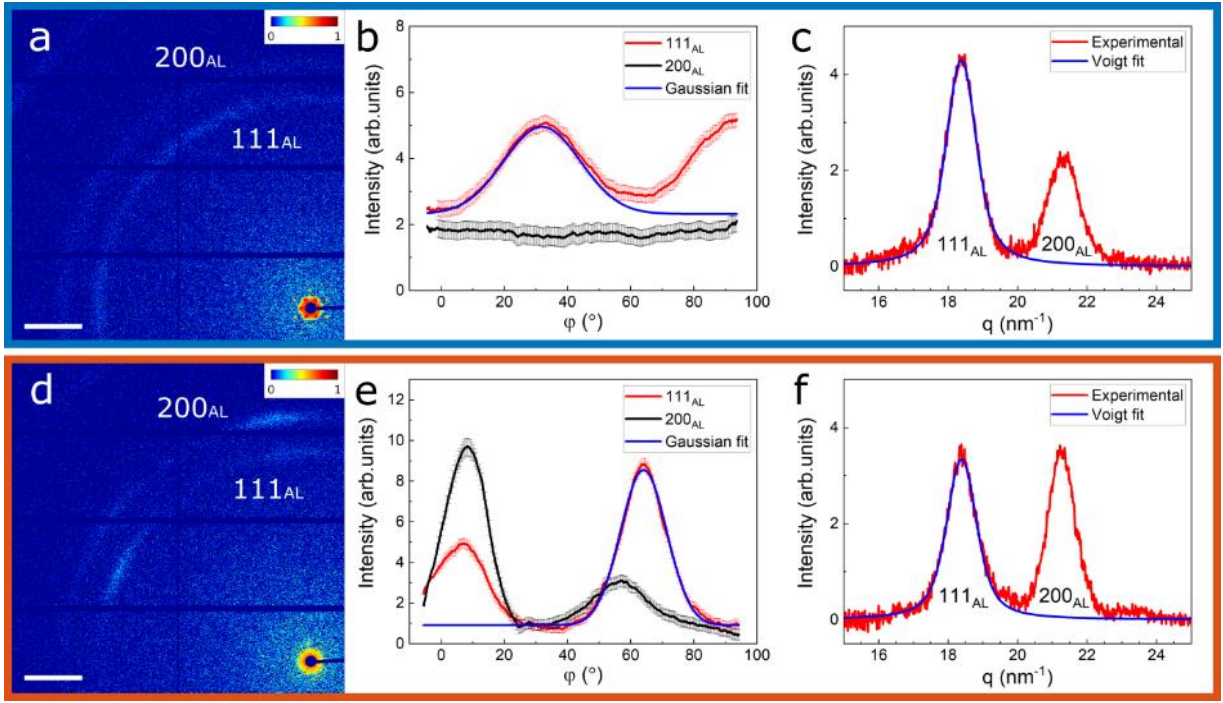

**Figure S9**: **WAXS study of mono- (a–c) and polycrystalline (d–f) channels**: (**a,d**) examples of a single WAXS diffraction pattern; (**b,e**) azimuthal and (**c,f**) radial profiles of the $111_{AL}$ and $200_{AL}$ reflections. A Voigt fit of the $111_{AL}$ Bragg peaks is shown in blue. The 0º angle for the azimuthal profiles corresponds to the top of the diffraction patterns. The positive angular direction is counterclockwise. Scale bars in (**a,d**) correspond to 5 nm$^{-1}$.

To quantitatively characterize angular disorder, we assumed that WAXS peak broadening is caused by two factors: Scherrer broadening due to the small size of the NCs and orientational disorder of the NCs in sites of the superlattice. The first factor affects both the radial and the azimuthal width of the peaks while the second one influences only the azimuthal width of the peaks. Assuming that these two factors are independent, we can estimate the value of the orientational disorder $\Delta\Phi$ from the relationship between the radial and azimuthal widths of the peak:

$$\delta_{az}^2 = \delta_{rad}^2 + \Delta\Phi^2 \qquad (S2)$$

where $\delta_{az}$ and $\delta_{rad}$ are the FWHM values of the peak angular size in the azimuthal and radial directions obtained from a Gaussian and Voigt fitting, respectively.

We analyzed the radial and azimuthal profiles for $111_{AL}$ and $200_{AL}$ reflections on a single pattern. Azimuthal profiles are shown in **Figure S9b,e** for mono- and polycrystalline channels, respectively. Azimuthal profiles for a monocrystalline channel contain relatively wide Bragg peaks of $111_{AL}$ reflections, and no peaks of $200_{AL}$ reflections are observed. Both profiles have anisotropic offsets corresponding to many disordered NCs besides the ordered ones. The $111_{AL}$ Bragg peak was fitted by a Gaussian profile giving the FWHM value of 24.5 ± 1.0°. Azimuthal profiles for a polycrystalline channel contain sharp Bragg peaks for both, $111_{AL}$ and $200_{AL}$ reflections. Fitting the $111_{AL}$ Bragg peak gives the FWHM value of 16.7 ± 0.5°.

Radial profiles of the $111_{AL}$ and $200_{AL}$ reflections are shown in **Figure S9c,f** for mono- and polycrystalline channels, respectively. Fitting of the $111_{AL}$ reflection by a Voigt profile gives the FWHM value of 1 nm$^{-1}$ (3.1°) for both types of channels. According to the Scherrer equation, this value corresponds to the 6.9 nm-sized coherently scattering domains. Taken the precision of the method, it is in good agreement with the NC size (~5.8 nm) and superlattice unit cell parameters studied in this work.

Considering the obtained values, the orientational disorder ($\Delta\Phi$) of the atomic lattices of NCs is roughly 24° for monocrystalline and 16° for polycrystalline channels. The value for polycrystalline channels is similar to the recently reported $\Delta\Phi$ for superlattices of oleic acid- and tetrathiafulvalene-linked PbS NCs.[5,6]

**Angular X-ray Cross-Correlation Analysis basics**

The Angular X-ray Cross-Correlation Analysis (AXCCA) method is widely used for the analysis of disordered or partially ordered systems such as colloids, liquid crystals, polymers etc. It is capable of providing insights into hidden symmetries, such as bond-orientational order or partial alignment of particles in the system. This method was also shown to be highly useful to study the angular correlations in mesocrystals.[5,7] While details and mathematical background on this method can be found elsewhere,[8] here we briefly summarize the main concepts.

AXCCA is based on the analysis of a two-point angular cross-correlation function (CCF) that can be calculated for each diffraction pattern as

$$C(q_{AL}, q_{SL}, \Delta) = \frac{1}{2\pi} \int_{-\pi}^{\pi} \tilde{I}(q_{AL}, \varphi) \tilde{I}(q_{SL}, \varphi + \Delta) d\varphi \quad (S3)$$

where $\tilde{I}(q, \varphi) = I(q, \varphi) - \langle I(q, \varphi) \rangle_\varphi$ and $I(q, \varphi)$ is an intensity value taken in the point with $(q, \varphi)$ polar coordinates in the detector plane and $\langle \cdots \rangle_\varphi$ denotes averaging over all azimuthal $\varphi$ angles. All values used in this definition are shown in **Figure S10**.

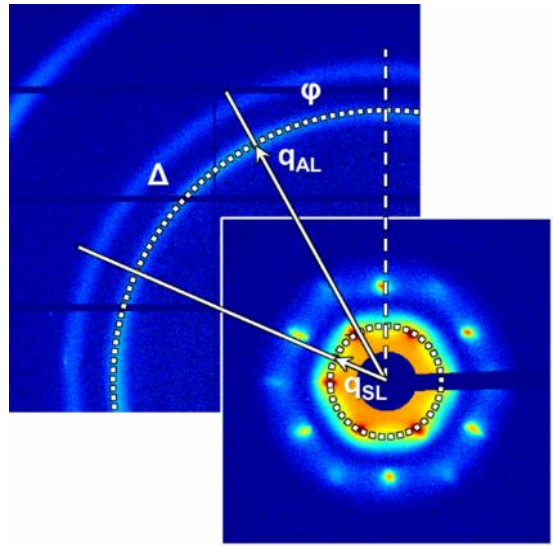

**Figure S10: Angular X-ray Cross-Correlation Analysis.** White arrows point to the Bragg reflections from the PbS AL and superlattice with momentum transfer values from the center of the pattern $q_{AL}$ and $q_{SL}$, respectively. The angle $\Delta$ between these Bragg peaks is shown. SAXS area is enlarged for better visibility.

Experimentally obtained diffraction patterns contain features that block the scattering signal, such as detector gaps, beamstop, beamstop holder, etc. In order to take their presence into account, we introduce into **Equation S3** the mask function

$$W(q, \varphi) = \begin{cases} 0, & \text{gaps, beamstop etc.} \\ 1, & \text{otherwise} \end{cases}. \tag{S4}$$

This yields the final form of the CCF as follows:

$$C(q_{SL}, q_{AL}, \Delta) = \int_{-\pi}^{\pi} \tilde{I}(q_{SL}, \varphi) W(q_{SL}, \varphi) \tilde{I}(q_{AL}, \varphi + \Delta) W(q_{AL}, \varphi + \Delta) d\varphi \tag{S5}$$

Taking the values of $q_{SL}$ and $q_{AL}$ indicated in the main text, we studied the correlations between reflections in the WAXS and SAXS areas. To obtain statistically meaningful data, CCFs were averaged over all diffraction patterns for each channel.

The CCF functions were simulated on the basis of the determined real-space structures. The Bragg peaks in both WAXS and SAXS areas were assumed to have Gaussian shapes in the angular direction, and the intensity on the corresponding ring was calculated as follows:

$$I(\varphi) = \sum_i \exp\left[-\frac{(\varphi - \varphi_i)^2}{\delta^2}\right] \exp\left[-\frac{(\theta - 2\theta_B)^2}{\delta^2}\right] \tag{S6}$$

where $\varphi_i$ is the azimuthal angular position of the $i$-th Bragg peaks in the SAXS/WAXS area, $\theta$ and $\theta_B$ are the angle between the detector plane and Ewald sphere and the Bragg angle for the considered reflection (used only for WAXS reflections). The angular sizes of the SAXS/WAXS peak $\delta$ were chosen to fit the experimental data.

The simulated CCFs were evaluated as

$$C_{sim}(q_{AL}, q_{SL}, \Delta) = \int_{-\pi}^{\pi} \tilde{I}_{SAXS}(q_{AL}, \varphi) \tilde{I}_{WAXS}(q_{SL}, \varphi + \Delta) \, d\varphi \tag{S7}$$

and then normalized by the maximum value of all CCFs for each channel.

**Angular X-ray Cross-Correlation Analysis results**

We calculated the CCFs for the first SAXS peaks ($\langle 1\bar{1}00\rangle_{AL}$ in the rhcp case) and the $111_{AL}$ or $200_{AL}$ WAXS peaks to reveal preferred angles between corresponding directions of the NCs and superlattice. Examples of the obtained CCFs for a monocrystalline rhcp channel are shown in **Figure S11c**.

For the monocrystalline channels, there are 6 peaks at 0°, ±60°, ±120° and 180º for both $C(q^{SL}_{1\bar{1}00}, q^{AL}_{111}, \Delta)$ and $C(q^{SL}_{1\bar{1}00}, q^{AL}_{200}, \Delta)$. The intensity ratio between them is ~8. These features correspond to the [0001] orientation of an rhcp superlattice, where the NCs are aligned as follows: $[111]_{AL}\|[0001]_{SL}$ and $[110]_{AL}\|[2\bar{1}\bar{1}0]_{SL}$. This configuration is shown in **Figure 3h** and confirmed by simulations (**Figure S11f**). It should be noted, that, as it follows from the WAXS analysis (see **Figure S9**), only a part of the NCs is aligned as shown.

However, a bcc superlattice oriented along $[111]_{SL}$ would give the same SAXS pattern. To verify the proposed rhcp structure, we simulated an CCFs for the bcc structure observed in polycrystalline channels. In such a bcc superlattice, all NCs directions are aligned with the corresponding directions of the superlattice. Thus, the orientation along $[111]_{SL}$ gives the same NCs orientation - $[111]_{AL}$. A real-space model of the considered superlattice is shown in **Figure S11g**. But the CCFs simulated for this structure do not correspond to the experimentally obtained ones (**Figure S11i**). To achieve the observed angular correlation between the superlattice and the NCs, the latter should be rotated by 30º around the $[111]_{AL}$ directions. The resulting structure is shown in **Figure S11j**. Indeed, such a structure gives the correct CCFs (shown in **Figure S11l**), but the assumed rotation breaks the symmetry of the entire structure and implies different NC orientations in the equivalent $\{111\}_{SL}$ planes.

Thus, we suppose that the monocrystalline channels form a symmetrical rhcp structure, where, for some fraction of the NCs, the $[111]_{AL}||[0001]_{SL}$ and $[110]_{AL}||[2\bar{1}\bar{1}0]_{SL}$ directions are collinear.

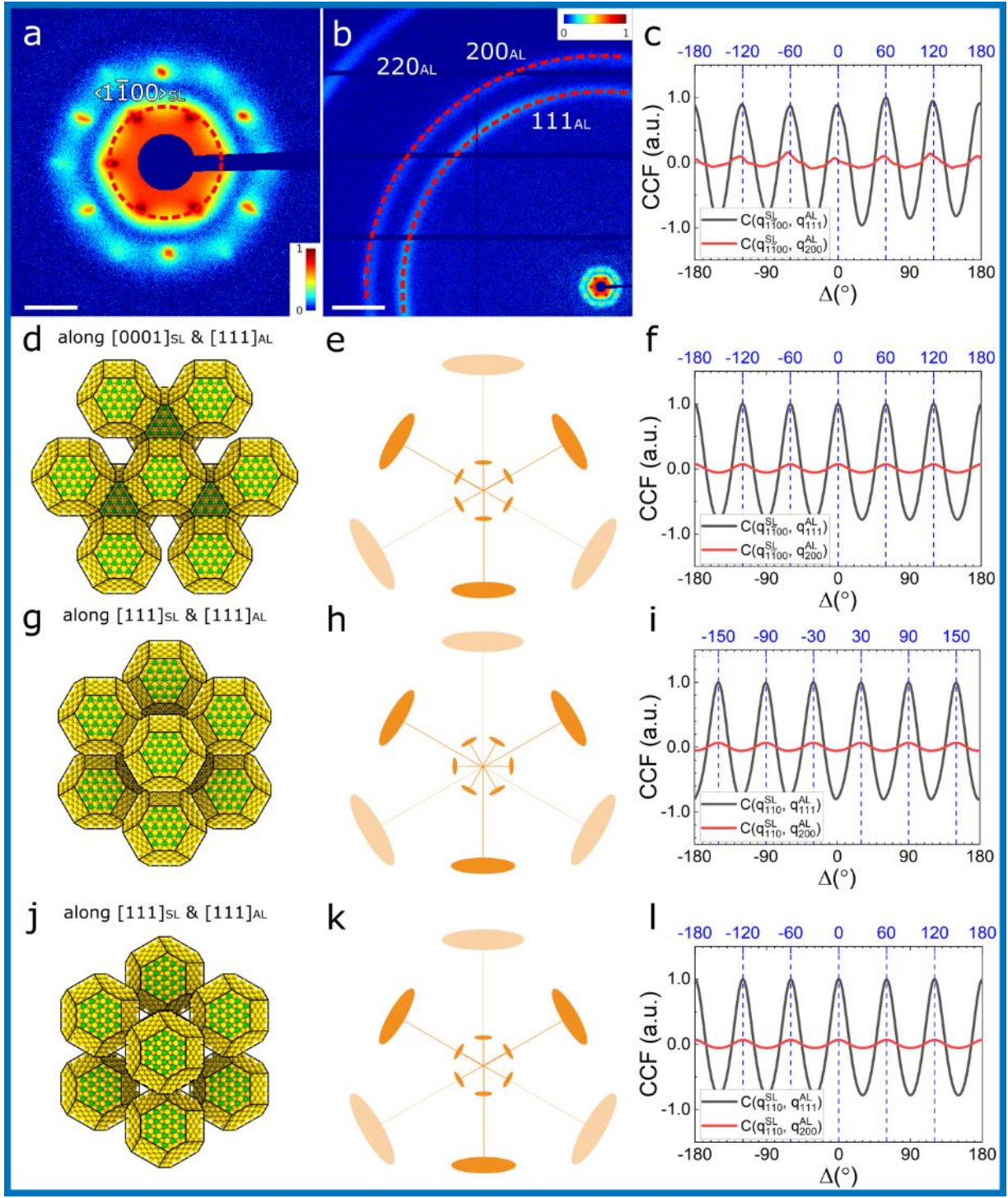

**Figure S11: AXCCA of the monocrystalline rhcp superlattice**: (**a**) An average SAXS pattern for the monocrystalline channel. The $1\bar{1}00_{SL}$ reflections used for AXCCA are shown with a red dashed line. The scale bar corresponds to 1 nm$^{-1}$. (**b**) An average WAXS pattern for a monocrystalline channel. The $111_{AL}$ and $200_{AL}$ reflections used for AXCCA are shown with red dashed lines. The scale bar corresponds to 5 nm$^{-1}$. (**c**) Calculated CCFs for the monocrystalline channel. (**d**) Proposed real-space rhcp superlattice model with $[111]_{AL}\|[0001]_{SL}$ and $[110]_{AL}\|[2\overline{1}\overline{1}0]_{SL}$; (**g**) considered bcc superlattice model with $[100]_{AL}\|[100]_{SL}$ and $[010]_{AL}\|[010]_{SL}$; (**j**) considered bcc superlattice model with $[111]_{AL}\|[111]_{SL}$ and $[1\overline{1}0]_{AL}\|[2\overline{1}\overline{1}]_{SL}$; (**e,h,k**) corresponding diffraction patterns (schematically, not to scale); (**f,i,l**) corresponding simulated CCFs for the first SAXS peaks ($\langle 1\overline{1}00\rangle_{SL}$ in the rhcp case (**f**) and $\langle 110\rangle_{SL}$ in the bcc case (**i,l**)) and the $111_{AL}$ or $200_{AL}$ WAXS peaks.

The CCFs for polycrystalline channels were calculated for the first SAXS peaks ($\langle 110\rangle_{AL}$ in the bcc case) and the $111_{AL}$ or $200_{AL}$ WAXS peaks. Examples of the obtained CCFs for a polycrystalline bcc channel are shown in **Figure S12c**. There are 4 peaks at $\pm 35°$, **±145°** for $C(q_{110}^{SL}, q_{111}^{AL}, \Delta)$ and two peaks at $\pm 90°$ for $C(q_{110}^{SL}, q_{200}^{AL}, \Delta)$. These features correspond to the $[110]_{SL}$ orientation of a bcc superlattice, where all crystallographic directions of the NCs are aligned with corresponding directions of the superlattice. This configuration is shown in **Figure S12d** and confirmed by simulations (see **Figure S12f**).

In the polycrystalline channels many different orientations are possible, but, according to our analysis, the $[110]_{SL}$ orientation is the primary one. To verify the latter, we simulated CCFs for other typical orientations. Examples for the same structures oriented along $[111]_{SL}$ (**Figure S12g**) and $[100]_{SL}$ (**Figure S12j**) are shown in **Figures S12i,l**, respectively. However, the features characteristic for these orientations are not observed in the experimental CCFs. Only a small fraction of the $[111]_{SL}$-oriented structures can contribute to the additional small peaks observed at $\pm 90°$ for the experimental $C(q_{110}^{SL}, q_{111}^{AL}, \Delta)$ (see **Figure S12c**). Other features can originate from the polycrystallinity (correlations between reflections from different grains are not compensated due to the lack of statics).

Thus, we assume the polycrystalline channels to have bcc superlattice structure, where all NCs are aligned with the corresponding superlattice directions (e.g. $[100]_{AL}\|[100]_{SL}$ and $[010]_{AL}\|[010]_{SL}$).

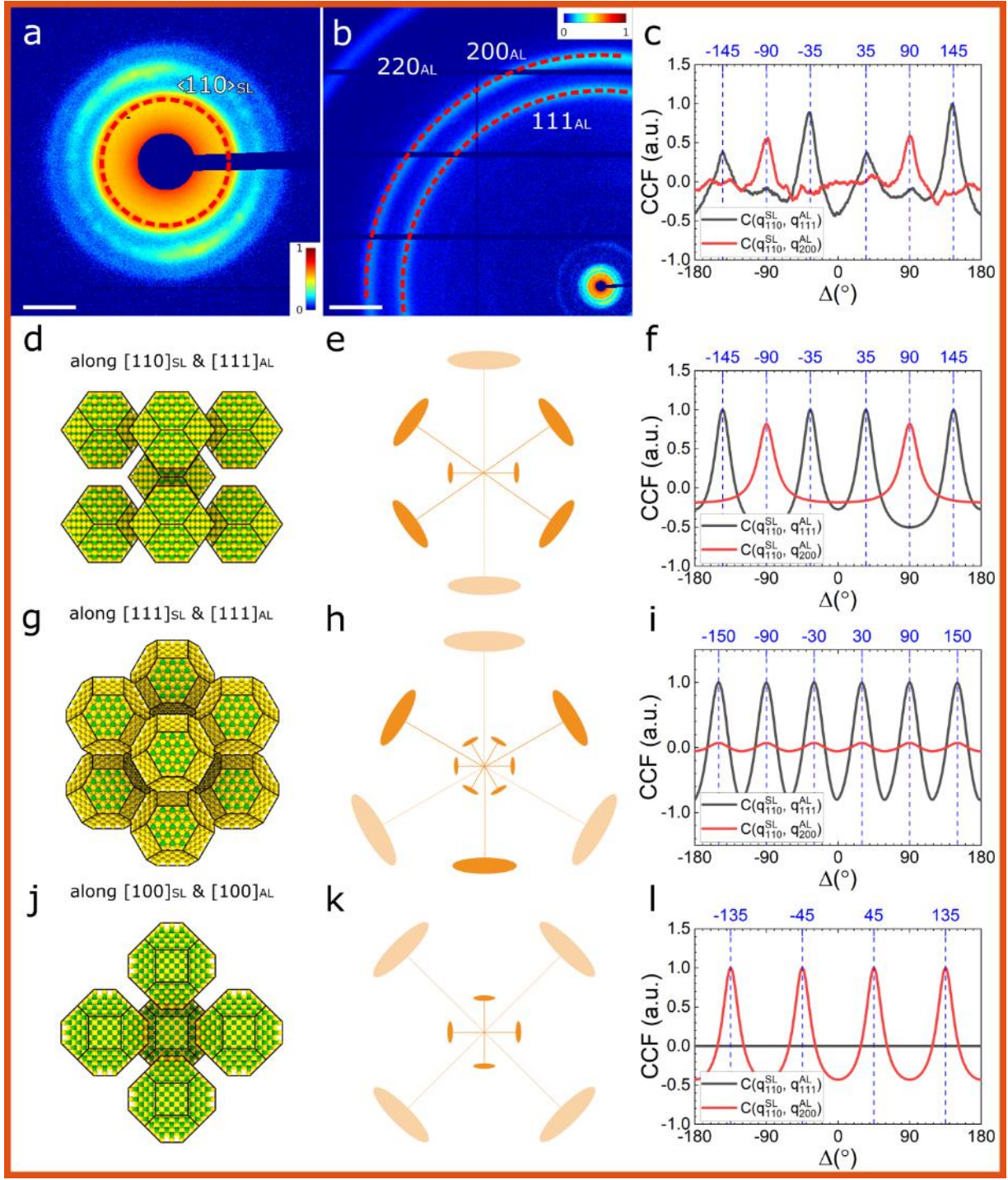

**Figure S12: AXCCA of the monocrystalline bcc superlattice**: (**a**) An average SAXS pattern for the polycrystalline channel. The $110_{SL}$ reflections used for AXCCA are shown with a red dashed line. The scale bar corresponds to 1 nm$^{-1}$. (**b**) An average WAXS pattern for a polycrystalline channel. The $111_{AL}$ and $200_{AL}$ reflections used for AXCCA are shown with red dashed lines. The scale bar corresponds to 5 nm$^{-1}$. (**c**) Calculated CCFs for the polycrystalline channel. (**d,g,j**) Proposed real-space bcc superlattice models with $[100]_{AL}\|[100]_{SL}$ and $[001]_{AL}\|[010]_{SL}$ oriented along (**a**) $[110]_{SL}$; (**d**) $[111]_{SL}$; (**g**) $[100]_{AL}$; (**e,h,k**) corresponding diffraction patterns (schematically, not to scale); (**f,i,l**) corresponding simulated CCFs for the first $\langle 110 \rangle_{SL}$ SAXS peaks and the $111_{AL}$ or $200_{AL}$ WAXS peaks.

**SEM investigation of the PbS NC superlattice types**

**Figure S13** shows scanning electron microscopy (SEM) micrographs of the different superlattice types, as described in the main text. SEM imaging corroborates the structural properties, revealed by the X-ray nano-diffraction.

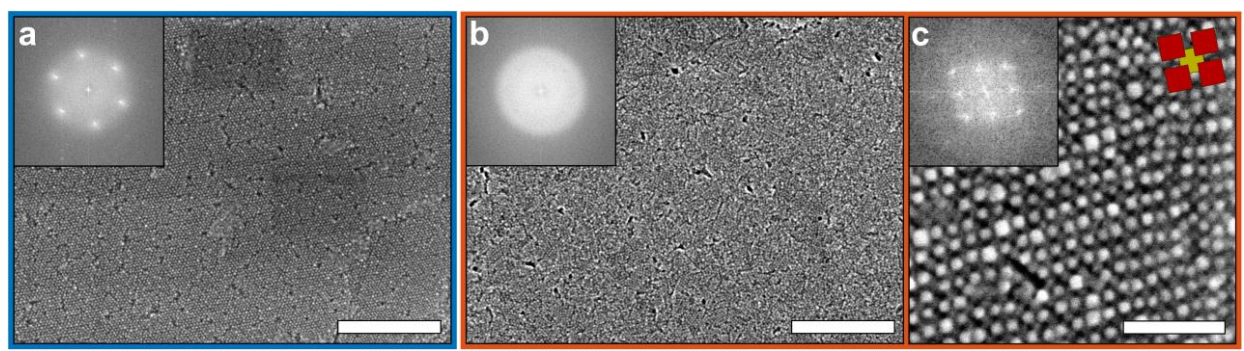

**Figure S13. SEM investigation of the two types of PbS NC superlattices.** (**a**) Monocrystalline superlattice with rhcp structure. Scale bar: 300 nm. (**b**) Polycrystalline superlattice with bcc structure. Scale bar: 300 nm. (**c**) High-resolution image of a bcc superlattice. Schematic drawing indicates alignment of individual NCs. Scale bar: 50 nm. The fast Fourier transformations (insets) support the polycrystalline and monocrystalline nature as well as the structure of the superlattices, as deduced from the X-ray nano-diffraction data.

**Semi-quantitative Raman-spectroscopy analysis**

The superlattices within microchannels exhibit strong characteristic Raman signals for Cu4APc (750 cm$^{-1}$ and 1,050–1,650 cm$^{-1}$),[9] as displayed in **Figure S14a**. Probing areas outside the superlattice stripes, these characteristic signals vanish, verifying the specific functionalization of the NCs with the organic π-system (see line scan in **Figure S14b**).

Raman-spectroscopy was performed on the corresponding Si/SiOx substrates, as the characteristic polyimide signals of the Kapton devices overlay with those of Cu4APc.

**Figure S14c** shows the Raman-spectroscopy analysis to semi-quantify the degree of ligand exchange and its correlation to superlattice type. Stronger Cu4APc signals are observed for polycrystalline bcc superlattices with a smaller lattice parameter (NND), compared to monocrystalline rhcp lattices. This supports the hypothesis of incomplete oleic acid (OA) ligand exchange in rhcp monocrystals. Raman signal from OA cannot be detected (**Figure S14e**).

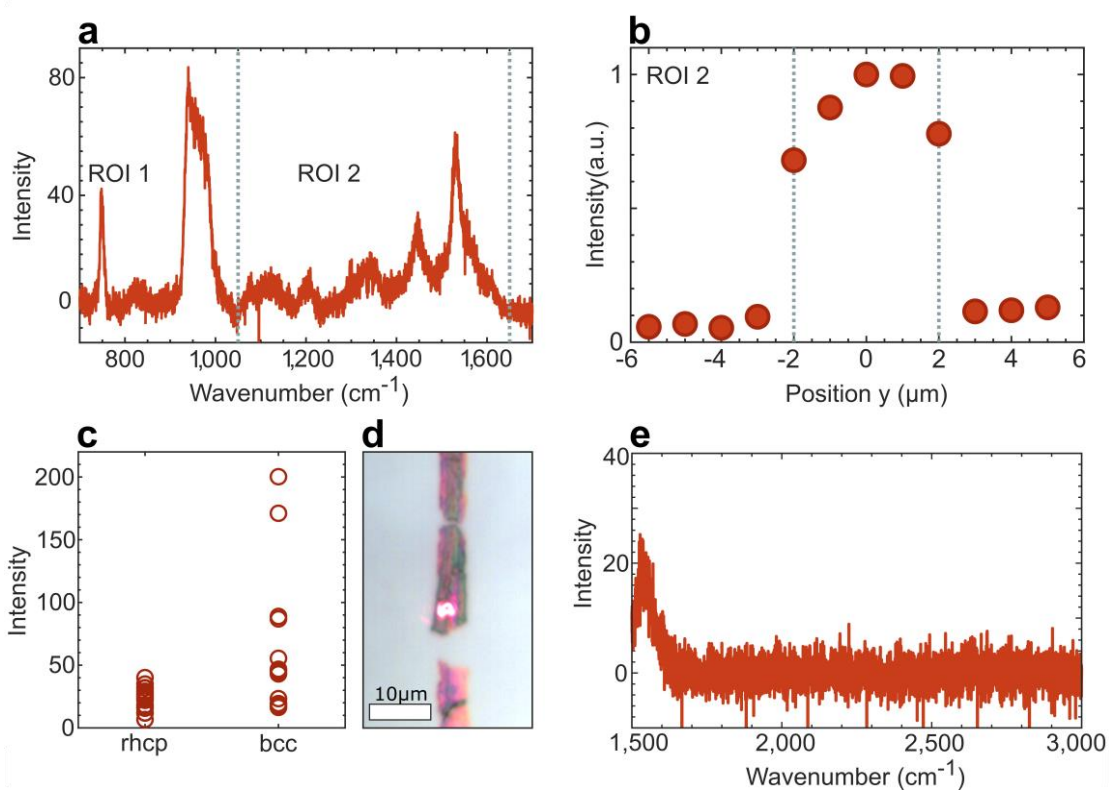

**Figure S14**: **Semi-quantitative Raman-spectroscopy analysis.** (**a**) Raman spectrum of a microcontact-printed superlattice stripe of PbS NCs functionalized with Cu4APc, showing characteristic signals of the ligand molecule at 750 cm$^{-1}$ and 1,050–1,650 cm$^{-1}$, indicated as ROI 1 and ROI 2, respectively. The signal at ~950 cm$^{-1}$ corresponds to the silicon background. (**b**) In a line scan across a stripe, signal from the characteristic fingerprint region

of Cu4APc (ROI 2) is only detected on the PbS NC stripe. (**c**) Comparison between the two superlattice types. The polycrystalline bcc superlattices with smaller lattice parameters (NND) exhibit stronger Raman signal from Cu4APc, supporting our hypothesis in the main text. (**d**) Typical PbS NC stripe on a Si/SiO$_x$ device with the 632 nm laser focus of the Raman-setup. (**e**) Raman-spectrum of a PbS NC stripe with monocrystalline hcp superlattice. The spectrum clearly lacks signal from oleic acid, which is supposed to appear at ~ 2,800 cm$^{-1}$.

**Qualitative investigation of anisotropic charge transport**

**Figure S15a** displays the graphical approach of identifying monocrystalline channels for which the parameters nearest-neighbor distance NND and thickness *h* are identical. Exactly four monocrystalline channels (2 pairs of 2) fulfil those requirements and can directly be compared. We observe higher conductivity *σ* for the channels with lower angle *α*.

For monocrystalline superlattices with similar crystalline order and lattice parameter (NND) but varying thickness, we normalize the measured electric conductivities by applying an empirical correction of the thickness dependence, for better comparability. This correction is obtained from fitting the thickness-dependent conductivity data of twenty individual microchannels (**Figure S15b–c**). **Figure S15b** shows the conductivity of microchannels as a function of superlattice thickness. All polycrystalline bcc superlattices exhibit almost identical lattice parameters and structural order. Thus, the difference in conductivity can only be attributed to different superlattice thicknesses. As described in the main text, the influence of *h* on *σ* is attributed to a fringing electric field along the height (sample normal), resulting in an inhomogeneous current flow. This effect should be identical for all superlattice types. The red line corresponds to the empirical fit of this conductivity-thickness dependence of polycrystalline channels. Applying this dependency to other microchannels allows us to normalize the conductivity of all channels to their corresponding thickness. This thickness-normalized conductivity is shown in **Figure S15c** as a function of nearest-neighbor distance. Here, monocrystalline rhcp superlattices with the same NND can be directly compared.

**Figure S15d** shows the thickness-normalized conductivity of monocrystalline rhcp microchannels as a function of azimuthal angle $\alpha$. For the sake of clarity, the conductivity of the respective more conductive superlattice is set to one. As a general result, the superlattices with lower value of $\alpha$ show higher conductivity. This finding strongly supports the hypothesis of anisotropic charge transport, as discussed in the main text.

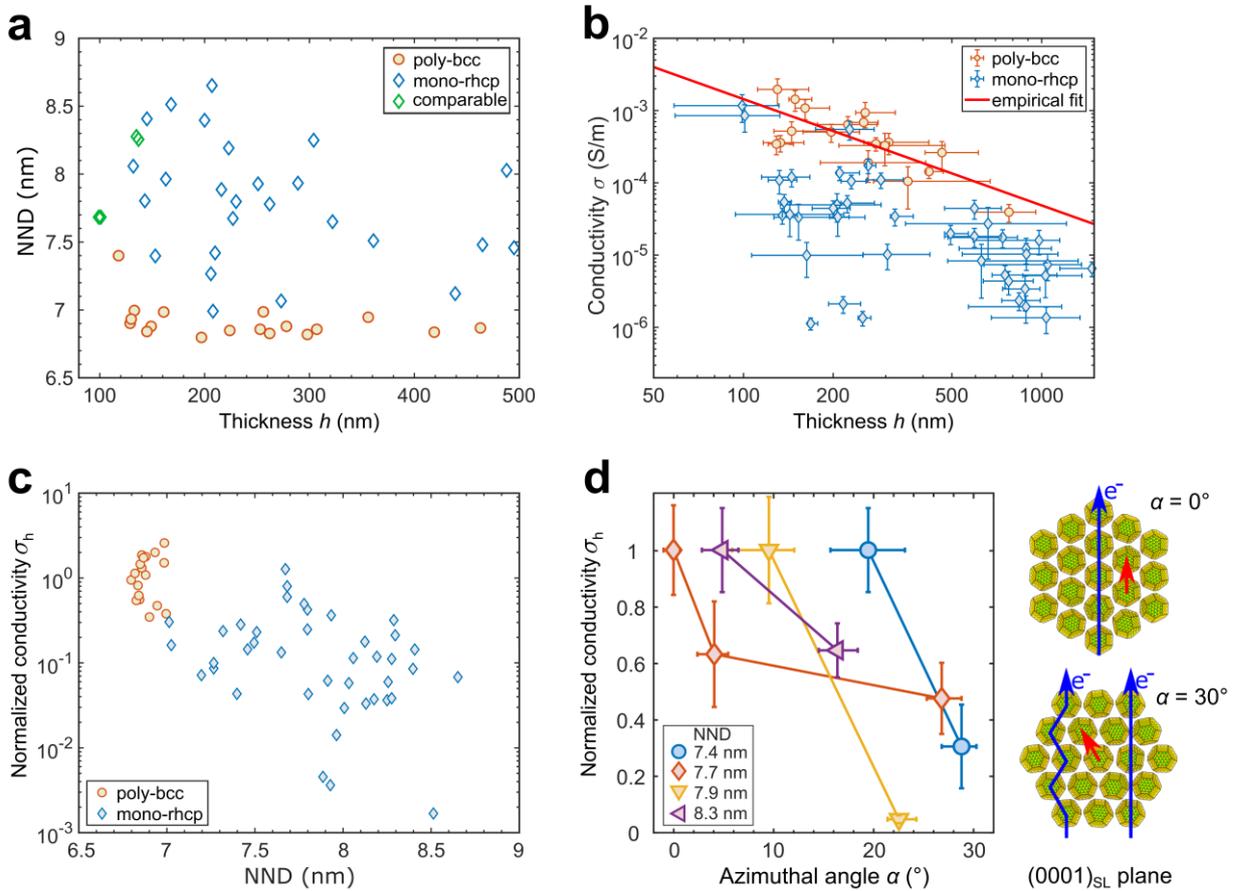

**Figure S15. Anisotropy of charge transport in monocrystalline NC superlattices.** (**a**) Graphical illustration of nearest-neighbor distance and thickness of investigated superlattices. Comparable monocrystalline channels with identical NND and *h* are highlighted in green. (**b**) Conductivity of microchannels as a function of superlattice thickness. Polycrystalline and monocrystalline microchannels are displayed. The red line corresponds to an empirical fit, which is applied for the thickness-normalization. (**c**) Conductivity normalized to the thickness as a function of nearest-neighbor distance NND. Now, channels with the same lattice parameter (NND) can be compared. (**d**) Comparison of rhcp monocrystals. Thickness-normalized conductivity of monocrystalline channels as a function of azimuthal

angle α. The color code indicates comparable superlattices with identical NND and the connecting lines are guides for the eye. For clarity, the conductivity of the respective more conductive superlattice is set to one. Inset: Schematic of the rhcp superlattice and the favoured hopping path for α = 0° (blue arrow) along the **d**$_{NN}$ direction (red arrow). For an in-plane offset (α = 30°), the larger hopping distance or the zig-zag path are detrimental to charge transport. Ligand spheres of NCs are omitted for clarity.

## On the origin of error bars

**Error of the nearest-neighbor distance:** Mean values are calculated from the multiple (1–3) Bragg peaks registered for one microchannel. Errors correspond to the standard deviation of those values.

**Error of thickness $h$:** The thickness $h$ is determined by SEM imaging under a tilted view of 85° with respect to the incoming beam (Si/SiO$_x$ devices) and AFM (Kapton devices). Mean values and ranges are determined by multiple measurements (SEM) or averaged height profiles (AFM).

**Error of conductivity $\sigma$:** Mean values of conductivity are calculated as described in the main text. Errors are calculated by Gaussian error propagation, given in **Equation S8**.

$$\frac{\Delta\sigma}{\sigma} = \sqrt{\left(\frac{\Delta G}{G}\right)^2 + \left(\frac{\Delta L}{L}\right)^2 + \left(\frac{\Delta W}{W}\right)^2 + \left(\frac{\Delta h}{h}\right)^2} \qquad (S8)$$

Here, $\Delta G$ corresponds to the error of $G$, determined by multiple $I$-$V$ curves of one microchannel. $\Delta L$ and $\Delta W$ correspond to the uncertainty of $L$ and $W$ determined by AFM. $\Delta h$ corresponds to the thickness variation measured by AFM.

**Error of azimuthal orientation:** The mean values correspond to the central peak position ($q_{1\bar{1}00}^{SL}$), and the error indicates the FWHM of the peak.

# Supporting References

1. Qin, D., Xia, Y. & Whitesides, G. M. Soft lithography for micro- and nanoscale patterning. *Nat. Protoc.* **5,** 491–502 (2010).

2. McCold, C. E., Fu, Q., Howe, J. Y. & Hihath, J. Conductance based characterization of structure and hopping site density in 2D molecule-nanoparticle arrays. *Nanoscale* **7,** 14937–14945 (2015).

3. André, A. *et al.* Structure, transport and photoconductance of PbS quantum dot monolayers functionalized with a copper phthalocyanine derivative. *Chem. Commun.* **53,** 1700–1703 (2017).

4. Liu, Y. *et al.* Dependence of carrier mobility on nanocrystal size and ligand length in PbSe nanocrystal solids. *Nano Lett.* **10,** 1960–1969 (2010).

5. Zaluzhnyy, I. A. *et al.* Quantifying angular correlations between the atomic lattice and the superlattice of nanocrystals assembled with directional linking. *Nano Lett.* **17,** 3511–3517 (2017).

6. Zaluzhnyy, I. A. *et al.* Angular X-ray cross-correlation analysis (AXCCA): basic concepts and recent applications to soft matter and nanomaterials. *Materials* **12,** 3464 (2019).

7. Mukharamova, N. *et al.* Revealing grain boundaries and defect formation in nanocrystal superlattices by nanodiffraction. *Small* **15,** 1904954 (2019).

8. Kurta, R. P., Altarelli, M. & Vartanyants, I. A. X-ray cross-correlation analysis of disordered ensembles of particles. Potentials and limitations. *Adv. Cond. Matter Phys.* **2013,** 1–15 (2013).

9. Tackley, D. R., Dent, G. & Ewen Smith, W. Phthalocyanines: structure and vibrations. *Phys. Chem. Chem. Phys.* **3,** 1419–1426 (2001).